%% file: jmelbourne_apj.tex
\documentclass{aastex}
\usepackage{emulateapj5}

\submitted{Submitted 22 August 2003; Accepted 29 October 2003 -- To appear in the February, 2004 AJ}

\lefthead{Melbourne \etal}
\righthead{Metal Abundances of KISS Galaxies II.}

\newcommand{\etal}{{\it et\thinspace al.}\ }
\newcommand{\NHA}{log$(\mbox{[\ion{N}{2}]}\lambda6583/\mbox{H}\alpha)$}
\newcommand{\nha}{[\ion{N}{2}]$\lambda6583/\mbox{H}\alpha\;$}
\newcommand{\OHB}{log$(\mbox{[\ion{O}{3}]}\lambda5007/\mbox{H}\beta)$}
\newcommand{\ohb}{$\mbox{[\ion{O}{3}]}\lambda5007/\mbox{H}\beta\;$}
\newcommand{\OIIHB}{log$(\mbox{[\ion{O}{2}]}\lambda\lambda3726,29/\mbox{H}\beta)$}

\begin{document}

\title{Metal Abundances of KISS Galaxies. II.\\
Nebular Abundances of Twelve Low-Luminosity Emission-Line Galaxies}

\author{Jason Melbourne \& Andrew Phillips}
\affil{UCO/Lick Observatory, UC Santa Cruz, Santa Cruz, CA 95064}
\email{jmel@ucolick.org, phillips@ucolick.org}

\author{John J. Salzer}
\affil{Astronomy Department, Wesleyan University, Middletown, CT 06459}
\email{slaz@astro.wesleyan.edu}

\author{Caryl Gronwall}
\affil{Department of Astronomy \& Astrophysics, Penn State University, University Park, PA 16802}
\email{caryl@astro.psu.edu}

\and

\author{Vicki L. Sarajedini}
\affil{Astronomy Department, University of Florida, Gainesville, FL 32611}
\email{vicki@astro.ufl.edu}

%\newpage

\begin{abstract}
We present follow-up spectra of 39 emission-line galaxies (ELGs) from the 
KPNO International Spectroscopic Survey (KISS).  Many targets were selected 
as potentially low metallicity systems based on their absolute B magnitudes 
and the metallicity-luminosity relation.  The spectra, obtained with the Lick 
3-m telescope, cover the full optical region from [\ion{O}{2}]$\lambda\lambda3726,29$
to beyond [\ion{S}{2}]$\lambda\lambda$6717,31 and include measurement of
[\ion{O}{3}]$\lambda$4363 in twelve objects.  The spectra are presented and tables 
of the strong line ratios are given.  For twelve high signal-to-noise ratio 
spectra, we determine abundance ratios of oxygen, nitrogen, neon, sulfur and 
argon.  We find these galaxies to be metal deficient with three systems 
approaching O/H of 1/25th solar.  We compare the abundance results from 
the temperature-based T$_e$ method to the results from
the strong-line $p_3$ method of Pilyguin (2000).    
\end{abstract}


\section{Introduction}
Star-forming emission-line galaxies (ELG's) are important probes of
galaxy evolution.  Spectra of ELG's reveal the abundances of heavy elements 
in the host galaxies (Searle and Sargent 1972; Izotov, Thuan \& Lipovetsky 1994; Izotov 
\& Thuan 1999).  The measured abundances are tracers of star-formation history 
and can be used to estimate star-formation rates (Kennicutt 1998, Contini et 
al. 2002), initial mass functions of the starbursts (Stasinska and Leitherer 
1996), and general trends in galactic evolution such as the the 
metallicity-luminosity relation (Skillman et al. 1989; Richer and McCall 1995; 
Melbourne and Salzer 2002, hereafter Paper I).

The study of metal abundances of star-forming ELG's has revealed
a class of extremely low-metallicity galaxies such as I
Zw 18 and SBS 0335-052, both with metallicities of roughly 1/50th
solar (Izotov et al. 1997).  In addition to these two extreme galaxies
are a handful of galaxies with metallicities roughly 1/25 solar
(Izotov et al. 1999).  These galaxies are all dwarf systems which follow 
a metallicity-luminosity trend whereby the less luminous galaxies tend 
to be less chemically evolved.   They place bounds on the primordial 
composition of the universe, an important observational constraint on 
Big Bang nucleosynthesis.  Izotov et al. (1994, 1997), among others, 
use these extremely metal-poor systems to infer a value for the 
primordial helium abundance.

This paper presents spectral follow-up observations of potentially 
metal-poor systems drawn from the KPNO International Spectroscopic 
Survey (KISS, Salzer et al 2000).   KISS identifies ELG candidates 
out to a redshift of $z=0.095$ by selecting objects that exhibit 
line emission in low-dispersion objective-prism spectra.  
The blue survey (Salzer at el. 2002) identifies objects which possess
a strong [\ion{O}{3}]$\lambda5007$ emission feature.  The red survey 
(Salzer et al. 2001, Gronwall et al. 2003) selects objects via the 
H$\alpha$ line.  The sample contains a wide range of star-forming
galaxies, such as starburst nucleus galaxies, HII galaxies, irregular
galaxies with significant star formation, and blue compact dwarfs.  
The survey is also sensitive to Seyfert galaxies and quasars with 
emission lines redshifted into the bandpass of the objective-prism
spectra.  The red survey, based on the H$\alpha$ selection criterion, 
remains sensitive to even the most metal-poor ELGs which may be
overlooked by surveys with detection criteria based on [\ion{O}{3}] 
lines, blue colors, or UV excess.   Therefore we expect that KISS 
contains several extremely metal-poor systems which will be interesting 
for abundance studies.  

The goals of the current paper are twofold.  First, we publish the
results of the follow-up spectroscopy obtained over two seasons at
the Lick Observatory, as part of the overall effort by members of the
KISS group to obtain slit-spectra of KISS ELG candidates.  Second, we
present the first results of our program to obtain metallicity estimates
for potentially low-abundance objects.  In Section 2 we discuss the 
galaxy sample, the observations, and the data-reduction methods.  In 
Section 3 we present the follow-up spectral data obtained at Lick and 
discuss the properties of the objects observed.  Section 4 presents 
a detailed analysis of the twelve objects with abundance-quality spectra.  
We measure properties of the nebular emission regions such as electron
temperature and electron density and calculate abundance ratios of
heavy elements with respect to hydrogen.  Section 5 discusses the
abundance results, placing them in the context of previous work and
exploring the effects of secondary metal production. In addition we 
we use the sample to examine the recent strong line oxygen abundance
methods of Pilyugin (2000, 2001).  Our results are summarized in
Section 6. 

\section{Observations and Data Reduction} 

\subsection{Sample Selection}

As mentioned above, the galaxies discussed in this paper were chosen 
from the KISS catalogs primarily as potential metal-poor systems.  
Observing lists were prepared for each Lick run using the criteria
spelled out below.  In addition, a number of objects were observed
for other purposes entirely.  For example, several faint KISS objects
with putative radio detections were observed in 2001 as part of a
different study (Van Duyne et al. 2003).  Hence, the KISS objects for 
which we present follow-up spectra represent a rather heterogeneous
sample, and include a smattering of AGNs and several luminous starburst
galaxies in addition to many low-luminosity star-forming systems.

Objects were chosen for follow-up as potential low-metallicity galaxies 
based on one of two criteria.  First, objects with absolute B magnitudes 
$M_B \gtrsim -16$ are found to be metal poor based on the 
metallicity-luminosity relation (see Paper I).  Absolute magnitudes can be 
derived from the KISS survey data (coarse objective-prism redshifts plus
B-band photometry, see Salzer et al. 2000).  Based on the published
survey data alone, we were able to generate lists of potentially
interesting objects using $M_B \gtrsim -16$ as our cutoff.  Second, 
follow-up spectra existed for several hundred KISS ELGs at the time 
these Lick observations were obtained.  These were nearly
all taken in ``quick look" mode, where each object was observed for
$\sim$10 minutes or less.  For strong-lined ELGs, this is sufficient to
obtain an accurate redshift and to assess the nature of the object,
even for faint candidates (B = 20 or fainter).  However, these short
exposures are not adequate for deriving accurate abundances.  Also,
most of the KISS follow-up spectra have been obtained using spectrographs
with no blue sensitivity, meaning that we often do not have data for 
several important lines, such as [\ion{O}{2}]$\lambda\lambda3726,29$.
Therefore, we re-observed several objects at Lick for which their 
low-metallicity nature had been revealed by certain strong emission-line 
ratios (e.g., \ohb and \nha) found in their existing follow-up spectra.  
We were particularly interested in objects for which [\ion{O}{3}]$\lambda4363$ 
was visible in the ``quick look" follow-up spectra. The 
[\ion{O}{3}]$\lambda4363$/[\ion{O}{3}]$\lambda4959+\lambda5007$ line
ratio is used to measure the nebular electron temperature.
Accurate nebular abundance determinations rely on this electron 
temperature measurement.

In general, these criteria allowed us to identify metal-poor starburst
galaxy candidates with a reasonable degree of success.   However, in a
few cases we discovered high-redshift Seyfert galaxies and quasars 
rather than low-luminosity starburst galaxies.  In some cases a strong 
[\ion{O}{3}]$\lambda$5007 is redshifted to the bandpass covered by the
red KISS objective-prism data and is misinterpreted as H$\alpha$ from 
a low-luminosity, nearby dwarf galaxy.  Unfortunately the only way to 
distinguish these cases is through follow-up spectroscopy, since the 
objective-prism data are too low resolution to allow one to distinguish 
between these two options.

\subsection{Observations}

We obtained spectra of KISS ELGs at the Lick 3-m telescope over the 
course of three observing runs: April 29, 2000; May 31 - June 1, 2000;
and April 27 - 29 2001.  The spectra were taken with the KAST double 
spectrograph which makes use of a dichroic beam splitter (D55) to obtain
the full optical spectrum in one observation.  We observed with a slit
width of 1.5 arcsec and used Grating 2 (600/7500) on the 
red side giving a reciprocal dispersion of 2.32 \AA/pixel and Grism 2
(600/4310) on the blue side giving a reciprocal dispersion of 1.85 \AA/pixel.  
The full spectral range provided by this setup was from 3500 \AA\ to 7800
\AA.  Exposure times ranged from 20 minutes to 80 minutes (see below),
with the longer observations taken in 20 minute increments and combined 
later.  Each night of observations included spectra of a Hg-Cd-Ne and Ne-Ar 
lamps to set the wavelength scale as well as several spectrophotometric 
standard stars for flux calibration. 

Our brighter targets were visible directly on the slit-viewing camera
available for the KAST spectrograph.  However, for fainter objects 
(roughly B $>$ 17) we could not see the sources clearly in the camera,
and were forced to use the following procedure.  We removed the grating
and opened the slit jaws, then took a short exposure image of the source.  
We then moved the telescope until the target was centered on the known
location of the narrowed slit.
While somewhat inefficient, this procedure allowed us to position even
extremely faint objects in the slit of the spectrograph.  The faintest
object we observed had a B magnitude of 21.24.  The slit orientation was 
fixed at a position angle of 90$^\circ$ (E-W), to minimize the effect 
of differential atmospheric refraction for our sample of targets, which
have declinations comparable to the latitude of the observatory.

The amount of integration time each object received varied depending on
the nature of the source.  For objects without previous follow-up spectra,
it was unclear whether obtaining a high S/N abundance-quality spectrum was
warranted.  After obtaining a preliminary exposure of $\sim$20 minutes, the 
resulting spectrum was evaluated in real time.  Objects that were found to contain
the auroral [\ion{O}{3}]$\lambda4363$ line of sufficient strength to yield
a high quality abundance were then observed for additional time.  In this 
way high signal-to-noise spectra were taken only for the most promising 
low-metallicity candidates.  Objects whose spectra did not contain the 
[\ion{O}{3}]$\lambda4363$ line were only observed for the single exposure.
In nearly all cases, these data are sufficient for determining fundamental 
properties about the galaxies, such as redshift, galaxy activity type, and 
values for the strong emission-line ratios. 

\subsection{Data Reduction}

The data reduction was carried out with the Image Reduction and Analysis 
Facility\footnote{IRAF is distributed by the National Optical Astronomy 
Observatories, which are operated by AURA, Inc.\ under cooperative agreement
with the National Science Foundation.} (IRAF).  Processing of the 2D spectral 
images followed standard 
methods.  All of the reduction steps mentioned below were carried out on
the red and blue spectral images independently.  The mean bias level was 
determined and subtracted from each image by the data acquisition software 
automatically.   A mean bias image was created by combining 15 zero-second 
exposures taken on each night.  This image was subtracted to correct the
science images for any possible 2D structure in the bias frames.  Flat
fielding was achieved using an average-combined quartz-lamp image that was 
corrected for any wavelength-dependent response.  

1D spectra were extracted using the IRAF APALL routine.  The extraction width
(i.e., distance along the slit) was selected independently for each source by
examination.  For most sources the emission region was unresolved spatially, 
so that the extraction width was limited to $\sim$5-6 arcsec.  Sky subtraction 
was also performed at this stage, with the sky spectrum being measured in 8-12 
arcsec wide regions on either side of the object window.  The Hg-Cd-Ne and 
Ne-Ar lamp spectra were used to assign a wavelength scale, and the spectra of 
the spectrophotometric standard stars were used to establish the flux scale.
The standard star data were also used to correct the spectra of our target
ELGs for telluric absorption.  This is important because a number of our ELGs 
are at redshifts where the [\ion{S}{2}] doublet falls in the strong B-band and 
would otherwise lead to a severe underestimate of the true line
flux.  The emission-line fluxes and equivalent widths were measured using the
SPLOT routine.  When multiple images of the same object were available, the 
spectra were reduced separately then combined into a single high S/N spectrum 
prior to the measurement stage.

The KAST spectrograph uses a dichroic beam splitter to break the spectrum into
red and blue sections.  In order to analyze the data it is necessary
to ensure that the red and blue regions are on a consistent flux scale.  To
achieve this, we made certain that the extraction regions used in APALL were
precisely the same for both the red and the blue sections.  We utilize the standard star
observations to measure any remaining small-scale shifts between the red and
blue sides.  To do this, we measured the flux in the continua of the standard
stars taken each night in 100 \AA\ bins from 5000 \AA\ to 5800 \AA. A plot
of flux vs. wavelength reveals any small shifts between the continuum
fluxes on either side of the dichroic break.  Since the image scales for the
red and blue sides are the same, and since the same spectral extraction regions
were used for both sides, the correction factors were expected to be small.
This turned out to be the case, with the derived correction factors for each
night of observations ranging from a few percent up to 10 - 12\%.  For
our instrumental setup, H$\alpha$ and H$\beta$ are nearly always on opposite
sides of the dichroic.  Therefore the main effect of the correction factor is to
modify the H$\alpha$/H$\beta$ ratio, and hence the measured value of $c_{H\beta}$
(see below).  Thus, this necessary procedure introduces an additional level
of uncertainty into the derivation of the nebular abundances (Section 4) in the
sense that it affects the reddening corrections.

For an initial estimate of the  internal reddening in each galaxy we calculate
$c_{H\beta}$ from the H$\alpha$/H$\beta$ line ratio.  $c_{H\beta}$ is then used 
to correct the measured line ratios for reddening, following the standard 
prescription (e.g., Osterbrock(1989):
\begin{equation}
\frac{I(\lambda)}{I(\mbox{H}\beta)}=
\frac{F(\lambda)}{F(\mbox{H}\beta)}exp[c_{H\beta}f(\lambda)]
\end{equation}
where f($\lambda$) is derived from studies of absorption in the Milky Way (using 
values taken from Rayo et al. 1982).  Estimates of $c_{H\beta}$ for each galaxy are 
given in Tables 1-3.  A more rigorous estimate of $c_{H\beta}$ is made for the 
abundance-quality spectra (Section 4.1, Table \ref{table:lineratioscor}).

%*****************************************************************************

\section{The Spectral Data}

%\placefigure{fig:spec}
\begin{figure*}[htp]
\epsfxsize=5.0in
\epsscale{1.9}
\plotone{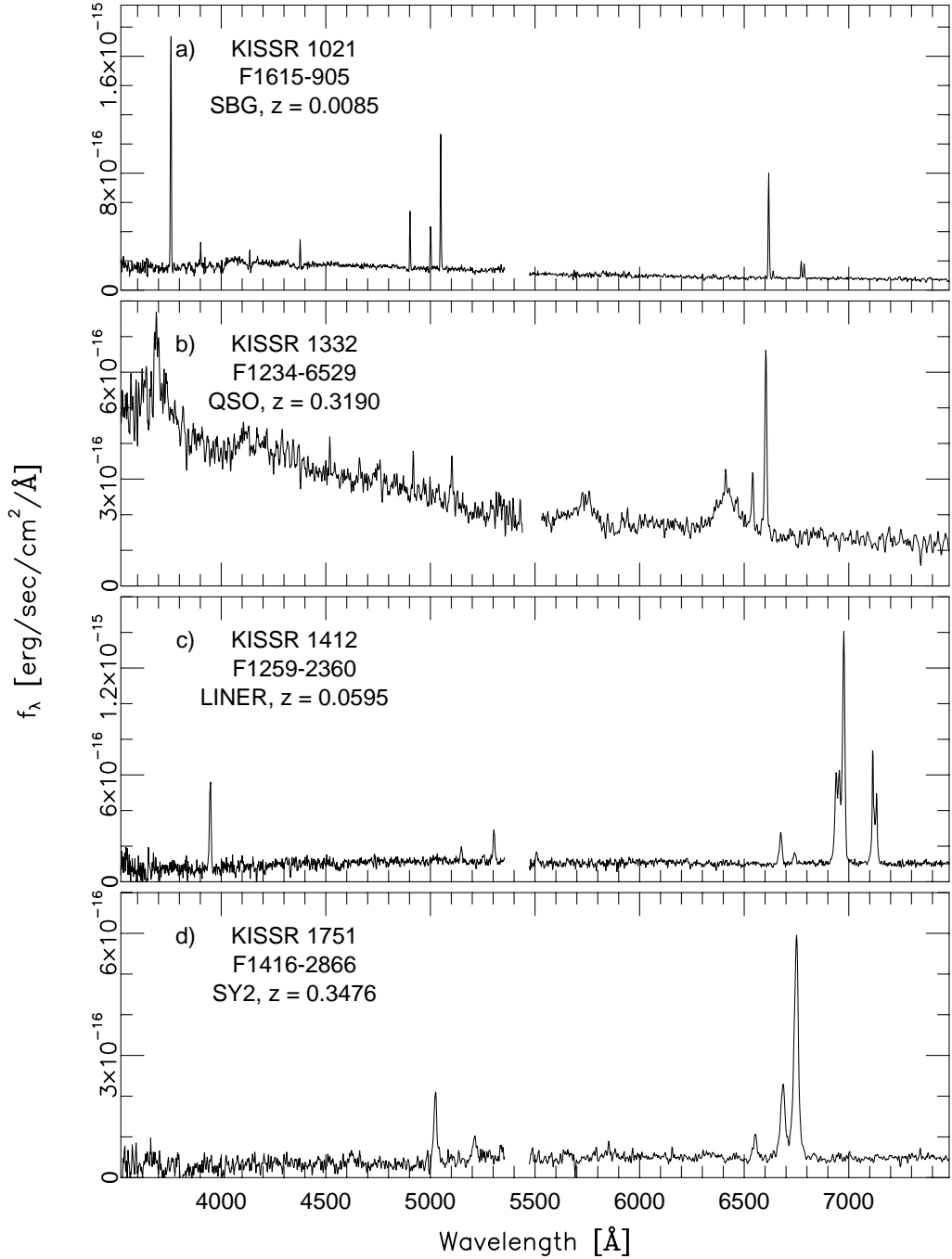}
\vspace{-0.3in}
\figcaption{Representative spectra of KISS objects obtained with the KAST spectrograph
  on the Lick Observatory 3-m telescope.  a) The starburst galaxy KISSR 1021.
  b) The QSO KISSR 1332, which is detected by KISS via the strong [\ion{O}{3}] line.
  c) The LINER KISSR 1412. d) The high redshift Seyfert 2 galaxy KISSR 1751, for
  which the [\ion{O}{3}] lines are shifted into the bandpass of the red KISS filter.
  The red and blue sides of these spectra have different dispersions.
  Therefore line ratios between lines from the two sides should not be
  compared using this Figure.  The flux scales on the red and blues
  sides were carefully calibrated (see Section 2.3).
  \label{fig:spec}}
\end{figure*}

The results of our spectroscopic observations of the Lick Sample are presented
in Tables \ref{table:lick30} - \ref{table:lickblue}.  The tables are organized by
the source KISS catalogs in which the targets are located.  Table \ref{table:lick30}
lists the results for KISS ELGS from the first red survey list (30$^\circ$ Red Survey;
Salzer et al. 2001), Table \ref{table:lick43} presents the spectral data for objects
from the second red survey list (43$^\circ$ Red Survey; Gronwall et al. 2003), and
Table \ref{table:lickblue} gives the data for the blue survey list (30$^\circ$ Blue 
Survey; Salzer et al. 2002).  All three tables have precisely the same format.
Column 1 gives the KISS number from the relevant catalog, while columns 2 \& 3 
list the survey field and ID number designation.  Column 4 indicates the observing
run during with the spectrum was obtained.  The values represent specific dates
from the various runs: 81 = 29 April 2000, 82 = 31 May 2000, 83 = 1 June 2000, 161 =
27 April 2001, and 163 = 29 April 2001.  Column 5 lists a coarse spectral quality
code.  Q = 1 refers to high quality spectra, with high S/N emission lines and 
reliable emission-line ratios.  Q = 2 is assigned to objects with lesser quality
spectra but still reliable line ratios.  Q = 3 refers to spectra where the data 
are of low quality, usually due to the faintness of the object and/or the weakness
of the emission lines.  Column 6 lists the redshift, obtained from the average of
the redshifts derived from all of the strong emission lines.  Typical formal
uncertainties for z are 0.00005 - 0.00010 (15 - 30 km/s).  Column 7 gives the decimal 
reddening coefficient c$_{H\beta}$, which is defined in equation 1.  In several cases the 
derived values for c$_{H\beta}$ are negative.  In virtually all instances where this
occurs, the formal error in c$_{H\beta}$ is such that the measured value is 
consistent with c$_{H\beta}$ = 0.00.  Whenever a negative c$_{H\beta}$ is measured,
we adopt c$_{H\beta}$ = 0.00 for the computation of reddening-corrected line ratios.

%\placetable{table:lick30}

%\placetable{table:lick43}

%\placetable{table:lickblue}

%\placefigure{fig:diagnostic}
\begin{figure*}[htp]
\epsfxsize=5.0in
\epsscale{1.2}
\vspace{-0.99in}
\plotone{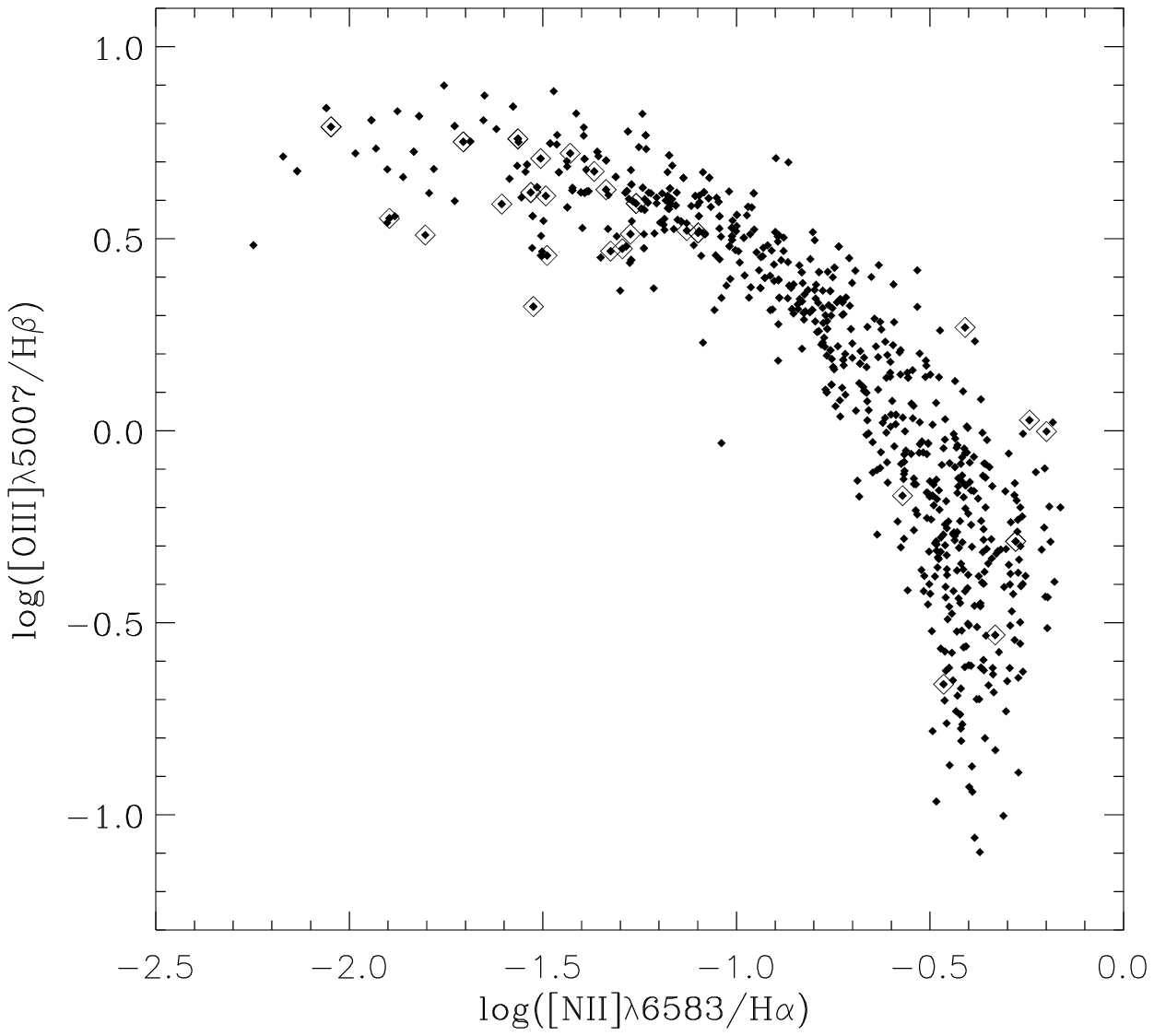}
\vspace{-0.5in}
\figcaption{A plot of \OHB\ as a function of \NHA\ for objects classified as
  starbursting types in the KISS database.  The filled symbols are objects that  
  have follow-up spectra in the KISS database of quality code 1 or 2. The open 
  diamonds represent the objects observed with the Lick 3m.  
  \label{fig:diagnostic}}
\end{figure*}

Columns 8 through 11 of the data tables present the observed equivalent
widths of [\ion{O}{2}]$\lambda\lambda$3726,29, H$\beta$, [\ion{O}{3}]$\lambda$5007,
and H$\alpha$.  Column 12 lists the H$\alpha$ flux, in units of 10$^{-16}$ ergs 
s$^{-1}$ cm$^{-2}$.  These values should be treated with caution, since the 
conditions during the observing runs were not always photometric, 
and the seeing was highly variable.
Columns 13 - 16 give the logarithms of the reddening-corrected line ratios for 
\OIIHB, \OHB, \NHA, and log([\ion{S}{2}]$\lambda\lambda$6717,31/H$\alpha$) 
for each galaxy.  If an equivalent width or line ratio measurement is not listed,
either the relevant line was not present in the spectrum (often due to low S/N),
or the line is redshifted out of the region covered by our spectra.  The latter
is true for lines like [\ion{N}{2}]$\lambda\lambda$6548,83, H$\alpha$, and
[\ion{S}{2}]$\lambda\lambda$6717,31 in the several objects with z $>$ 0.3.
Finally, Column 17 indicates the activity type for each object: SBG = Starburst 
Galaxy (i.e., any type of star-forming ELG), SY2 = Seyfert 2, LIN = low-ionization 
nuclear emission region
(LINER), QSO = quasi-stellar object, Gal = non-emission-line galaxy, Star = Galactic
star.  The latter two categories represent objects that are not actually extragalactic
emission-line sources, and hence are objects mistakenly identified by the survey as
ELGs.  To date, follow-up spectroscopy has shown that $\sim8-9$\% of the objects
cataloged in KISS are actually non-ELGs.

Characteristic spectra of objects observed with the Lick 3-m telescope are presented 
in Figure \ref{fig:spec}.  These were selected to illustrate the range of objects
identified by KISS.  Shown are the spectra of a low redshift star-forming galaxy (Figure 
\ref{fig:spec}a), a moderate redshift QSO (Figure \ref{fig:spec}b), a LINER (Figure 
\ref{fig:spec}c), and a moderate redshift Seyfert 2 galaxy (Figure \ref{fig:spec}d).
Both the QSO and the Seyfert 2 are examples of objects selected from the red survey
(H$\alpha$-selected) where another emission line ([\ion{O}{3}]$\lambda5007$ in both
cases) is redshifted into the low-z H$\alpha$ region.  About 2\% of the KISS galaxies with
follow-up spectra fall in this category; most are AGNs.  A more complete discussion of 
the AGN population found in the KISS catalogs is given in Gronwall, Sarajedini \& Salzer
(2002) and Stevenson et al. (2003).  The spectra of the 12 objects for which nebular 
abundances are derived are shown in the following section.

In Figure \ref{fig:diagnostic} we present a diagnostic diagram (e.g., Baldwin, Phillips
\& Terlevich 1981; Veilleux \& Osterbrock 1987) plotting \OHB\  against \NHA\  for all 
KISS objects with follow-up spectra of quality code Q = 1 or 2 which are classified as
star-forming types (points).   The galaxies observed at Lick are shown as diamonds.  Different 
types of star-forming galaxies are located in different regions of the plot.  Low metallicity 
starbursts are found in the upper left.  High metallicity starbursts are located in the 
lower right.  AGN and LINERS are found in the upper right portion of the diagram.
Only one of the AGNs observed at Lick has the relevant emission-line ratios necessary to
be plotted in Figure \ref{fig:diagnostic} (KISSR 1412, a LINER).  It falls off the diagram,
to the right.  All of the other AGNs are high redshift objects that lack the 
[\ion{N}{2}]$\lambda$6583/H$\alpha$ ratio.
Note that most of the objects in the Lick sample are located in the upper left section of the
diagram, suggesting that they are metal-poor objects.  This is exactly what one would
expect given the selection criteria.

Of the 39 objects in our sample,  29 are classified as starburst/star-forming
galaxies of some type, three are Seyfert 2s, one is a LINER, one is a QSO.  Five objects
are non-ELGs: two are Galactic stars and three are galaxies with no emission lines.
A more complete discussion of the spectroscopic properties of these galaxies will be 
presented in a future paper summarizing the spectroscopic follow-up of the entire KISS 
sample.

\section{Metal Abundances}    

Twelve galaxies in our sample were observed with a high enough
signal-to-noise ratio to detect [\ion{O}{3}]$\lambda4363$. 
The  [\ion{O}{3}]$\lambda4363$ line provides a measure of the 
electron temperature in the star-forming regions of these
galaxies.  Using the temperature in conjunction with the measured line
ratios we calculate accurate metal abundances.

%\placetable{table:galprop}

General properties of the sample are given in Table \ref{table:galprop}.  
Column 1 gives the KISSR/KISSB number of the object.  Column 2 shows the 
apparent $B$ magnitudes which range from 15.6 to 19.9.  The median m$_B$ 
is 17.6, and the sample contains three galaxies with $m_B > 19.5$,
significantly fainter than most objects found in traditional
nebular abundance studies.  Column 3 gives the $B-V$ color of each galaxy 
corrected for Galactic reddening.  Column 4 shows the absolute $B$
magnitude of each galaxy.  $M_B$ varies from -12.5 to -17.5, and thus
tends to be on the fainter end of the KISS ELG luminosity distribution.
This is not surprising since we were targeting objects that were
expected to have low metallicity.    Column 5 gives the recessional
velocity in km/sec, and Column 6 gives the calculated metallicity of
the object (see Sections 4.1 - 4.3).  Overall the sample galaxies tend to 
be fainter and more distant than the galaxies studied by previous groups 
doing nebular abundances.  While this makes it more difficult to do 
accurate abundance work, we find that reasonable results are possible. 
Not surprisingly, the formal errors in our final abundance estimates 
tend to be somewhat higher than those obtained in many of these previous
studies.  However, the precision of our results is still adequate for
most applications to the statistical study of chemical evolution in
galaxies.

The following section details the metal abundance determinations.  We will 
refer to the temperature-based abundance determination as the T$_e$ method.

\subsection{The Spectra and Line Ratios}

\begin{figure*}[htp]
\epsfxsize=5.0in
\epsscale{1.9}
\plotone{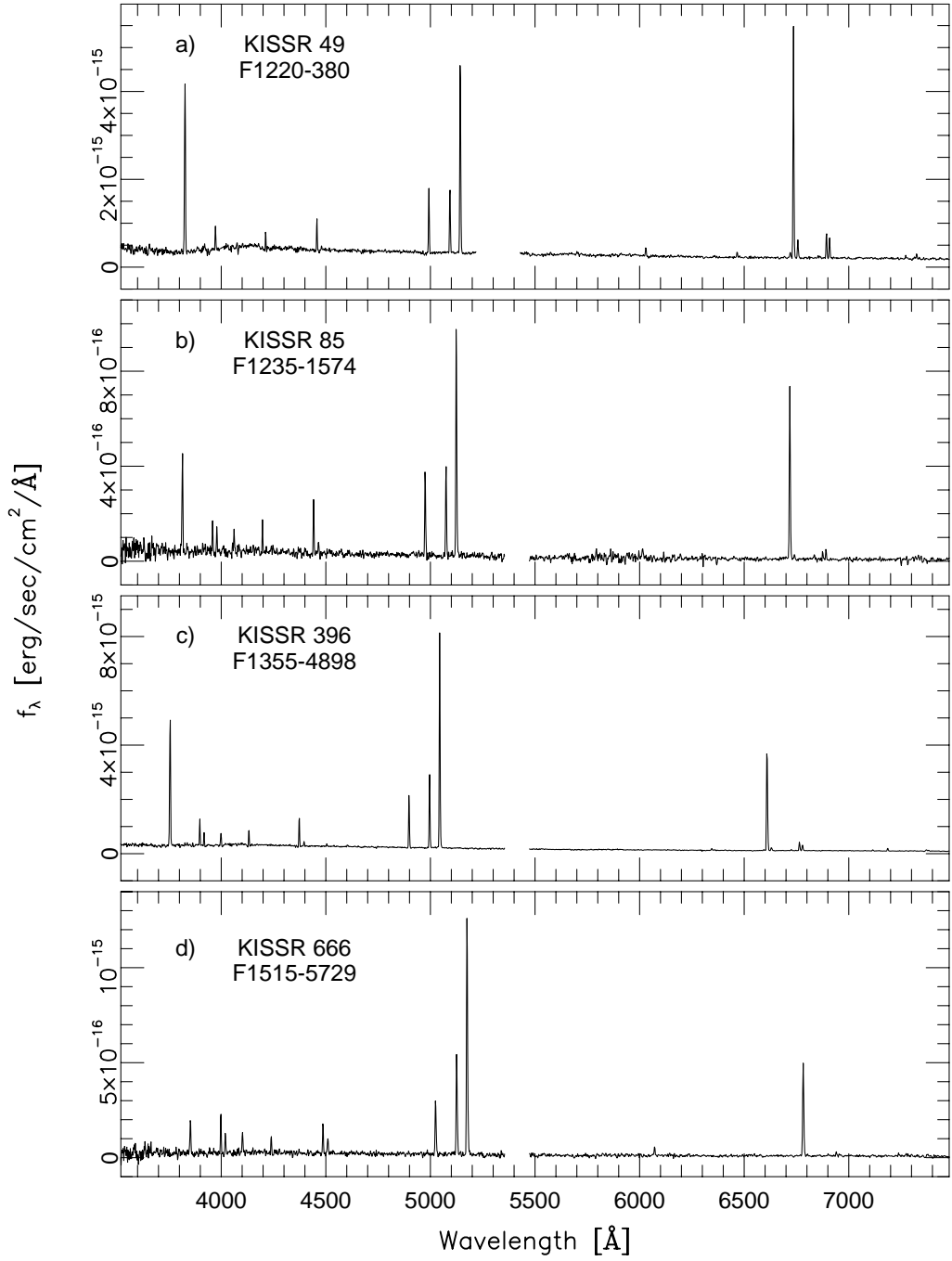}
\vspace{-0.3in}
\figcaption{Plots of the 12 abundance-quality spectra obtained with the
  Lick Observatory 3-m telescope.   As with Figure \ref{fig:spec}, 
  the red and blue sides of these spectra have different dispersions.
  Therefore line ratios between lines from the two sides should not be
  compared using this Figure. Specifically, the apparent H$\alpha$/H$\beta$
  line ratios will not be correct in this Figure.
  \label{fig:abunspec}}
\end{figure*}

The spectra of our 12 abundance-quality objects are presented 
in Figure \ref{fig:abunspec}.  These sources are characterized by low 
continuum emission with very strong emission features. These spectra 
generally have a high \OHB\  ratio and a low \NHA\ ratio indicative of  
low metal abundance and high excitation star-forming regions. The spectra
also contain detections of sulfur, neon, argon, and helium lines.  Measurements 
are made of all the lines with equivalent width exceeding 1 \AA.
Raw and reddening-corrected line ratios are presented in Tables
\ref{table:lineratios} and \ref{table:lineratioscor}, respectively.

%\placefigure{fig:abunspec}

\begin{figure*}[htp]
\epsfxsize=5.0in
\epsscale{1.9}
\plotone{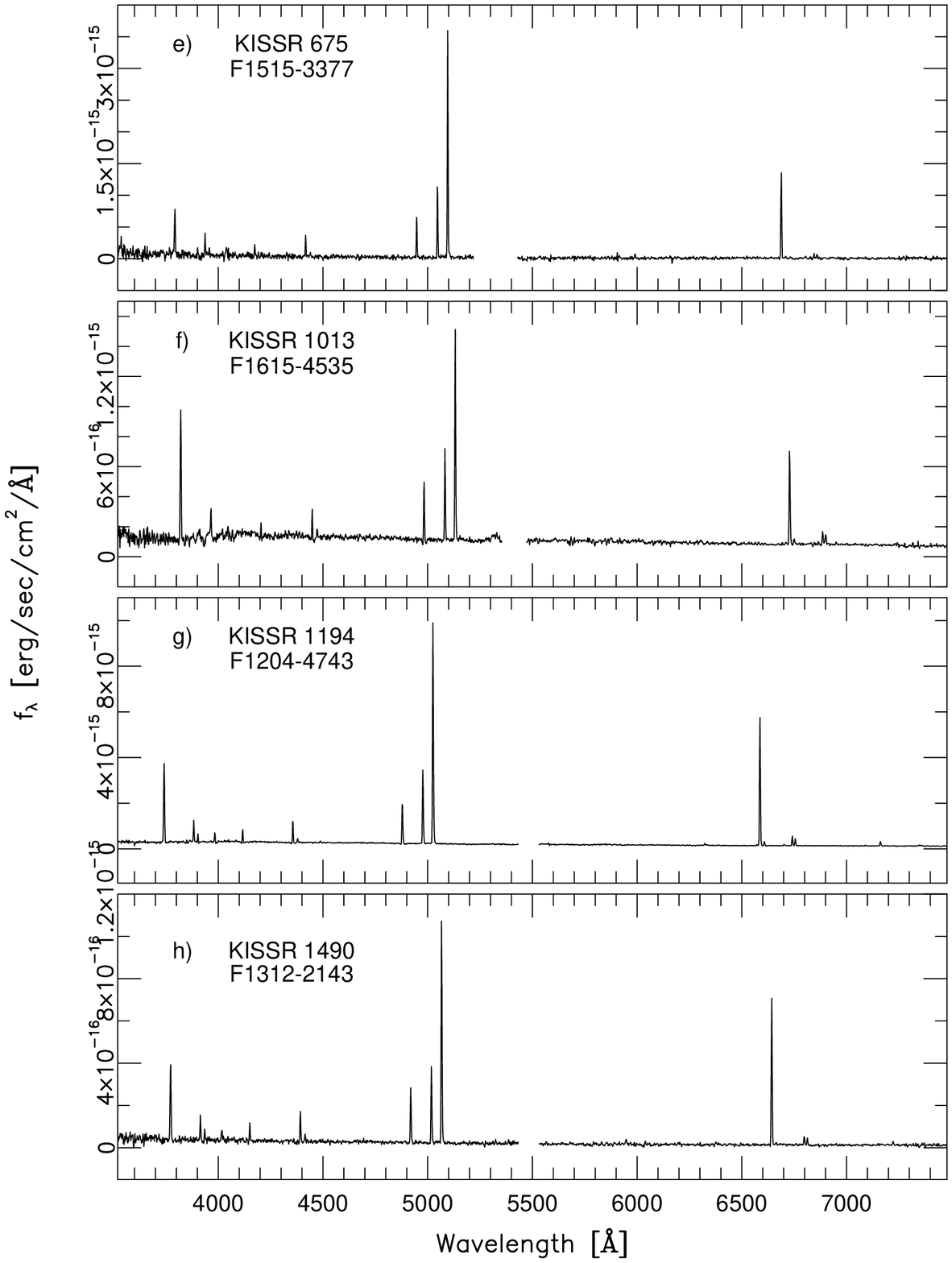}
  \label{fig:starburstb}
\end{figure*}

\begin{figure*}[htp]
\epsfxsize=5.0in
\epsscale{1.9}
\plotone{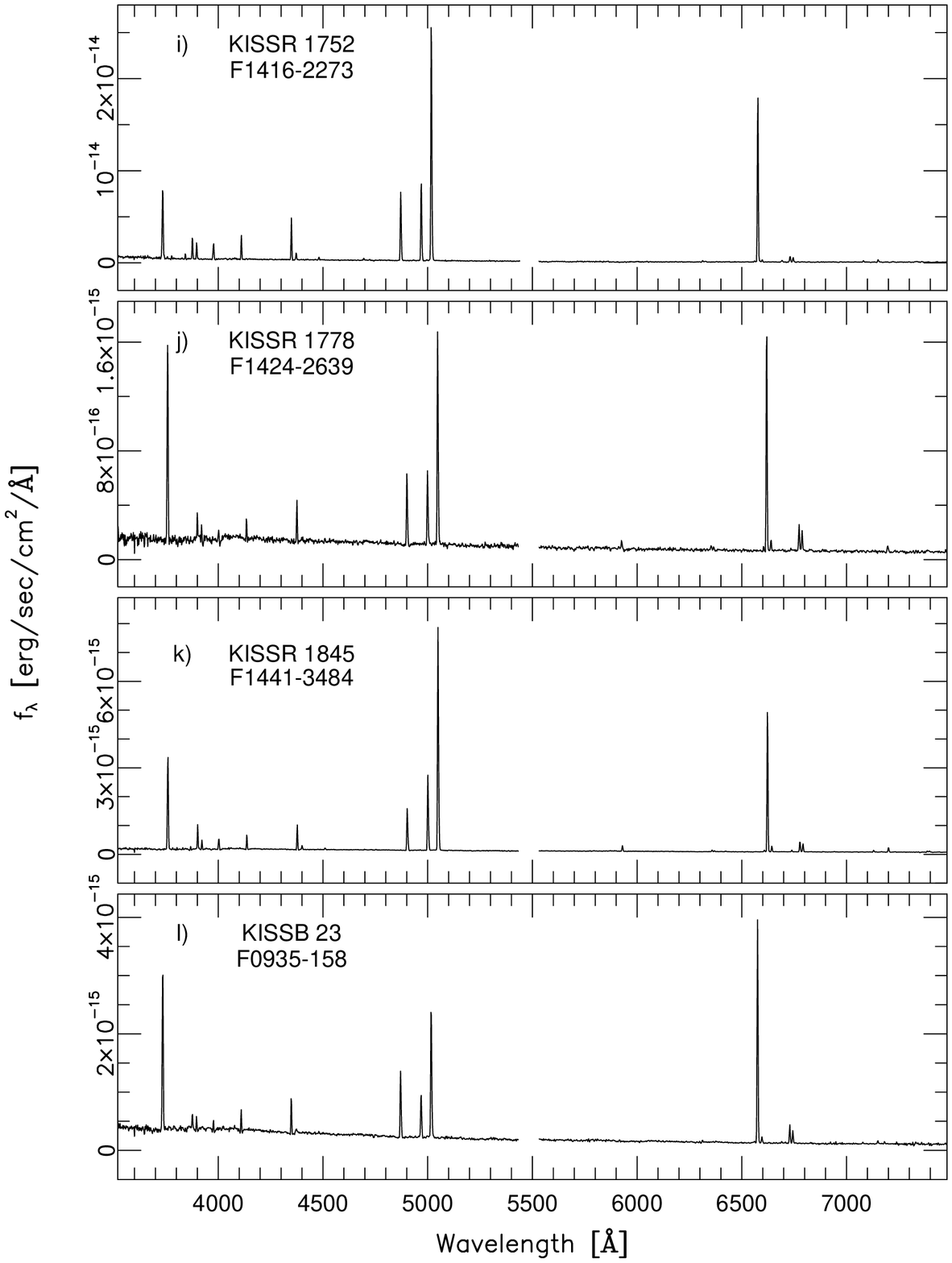}
  \label{fig:starburstc}
\end{figure*}

%\placetable{table:lineratios}

%\placetable{table:lineratioscor}

The $c_{H\beta}$ values used for the reddening correction were
determined from a  simultaneous fit to the reddening and absorption 
in the Balmer lines, using all available Balmer line ratios. The stellar 
absorption, usually not exceeding 3 \AA\ of equivalent width, is corrected for,
and a value for $c_{H\beta}$ is obtained for each galaxy by using the
average of the values for the three highest-order Balmer line ratios.
Because KAST is a dual channel spectrograph there is the added
complication of putting the red and blue sides on the same flux
scale.  In certain cases this caused problems with determining an
accurate value for $c_{H\beta}$.  However, we posses additional spectra of these
objects that cover the full optical range in one image.  We use these
additional spectra to confirm and in some cases revise the
H$\alpha$/H$\beta$ line ratio used to infer the reddening corrections.
The adopted values of   $c_{H\beta}$ are given in Table \ref{table:lineratioscor}.
(Note: these values are different than the estimated $c_{H\beta}$
values given in Tables 1-3.)

Four of the twelve galaxies have been observed spectroscopically by other
groups.  KISSR 49, also known as CG 177, was observed by Salzer et al. (1995) 
as part of a follow-up of Case ELGs.  The uncorrected[\ion{O}{2}]/H$\beta$ 
and [\ion{O}{3}]/H$\beta$ line ratios from the two studies match within the 
errors.  KISSR 396, also known as Was 81, was first observed by Wasilewski 
(1983) using a SIT vidicon spectrograph. He measured   [\ion{O}{3}]/H$\beta 
= 2.8$, while we find [\ion{O}{3}]/H$\beta = 4.1 \pm 0.1$.  A confirming 
spectrum from the Michigan-Dartmouth-MIT (MDM) 2.4-m telescope (Wegner et 
al. 2003) agrees with all the major Lick line ratios to within the 
errors, indicating that the more recent results are reliable.  KISSR 1752, 
the extremely metal-poor object SBS 1415+437 was observed by Thuan et al. 
(1999). The high signal-to-noise-ratio observations taken by Thuan et al. 
are equivalent to the Lick results to within errors for all the major indices. 
Finally KISSR 1845, also known as CG 903 and HS 1440+4302 was observed by 
Popescu \& Hopp (2000) in a survey of dwarf galaxies in voids and by Ugryumov et 
al. (1998, 1999).  In this case we find agreement with Popescu et al. to within 
the errors for the strong [\ion{O}{3}]/H$\beta$ line ratio.  However, they find
[\ion{O}{2}]/H$\beta = 5.30$ while we find [\ion{O}{2}]/H$\beta =
2.43$.  Some of this difference can be accounted for by discrepant
$c_{H\beta}$ measurements.  We found $c_{H\beta} = 0.113$ while they
found $c_{H\beta} = 0.336$.  The additional discrepancy is probably
due to the flux calibration at the blue end of the spectrum which tends to be
difficult to do consistently, or to the significant difference in slit width
used by the the two studies (1.5\arcsec\ for our study, 4\arcsec\ for 
Popescu).  In summary, our [\ion{O}{3}]/H$\beta$ line ratios compare
well with previous studies while in one case the 
[\ion{O}{2}]/H$\beta$ ratios vary significantly.
 
\subsection{Electron Density and Temperature}

We assume a two zone model for the star-forming nebula, a 
medium-temperature zone 
where oxygen tends to be doubly ionized and and low-temperature zone
where oxygen is singly ionized.  Within the radius of the
low-temperature zone hydrogen is ionized and beyond this zone
hydrogen is assumed to be neutral.

Ideally we would like to know the electron
density and temperature in each zone.  However, the data available
only allow for an electron density measurement in the low-temperature
zone, and an electron  temperature measurement in the 
medium-temperature zone.  We use the [\ion{S}{2}]$\lambda$6716/$\lambda$6731 
(Izotov et al. 1994) ratio to determine the electron 
density, which in all measurable cases is roughly 100
e$^-$cm$^{-3}$.  In several cases the sulfur lines are too noisy to
accurately determine the 
electron density.  In half of these cases the 
[\ion{S}{2}]$\lambda$6716/$\lambda$6731 doublet falls within the atmospheric 
B-band.   For galaxies where electron density was not calculated, 
we assume a density of 100 e$^-$cm$^{-3}$. 
In all cases we assume that the density does not vary
significantly from zone to zone.  
The electron temperature in the
medium temperature zone is given by 
the oxygen line ratio [\ion{O}{3}] ($\lambda4959 +
\lambda5007$)/$\lambda$4363 (Izotov et al. 1994).  Calculations of
both the electron 
density and electron temperature are carried out using the IRAF
NEBULAR package (Shaw \& Dufour 1995), which makes use of the
latest collision strengths and radiative transition probabilities. 

We estimate the temperature in the low-temperature zone by using the algorithm
presented in Skillman et al. (1994) based on nebular models of Stasinska (1990)
\begin{equation}
\label{O2temp}
t_e(OII) = 2((t_e(OIII)^{-1} + 0.8)^{-1},
\end{equation}
where t's are temperatures measured in units of $10^4$ K.  The measured electron 
densities and temperatures are presented in Table \ref{table:abund}.

Metal abundance measurements depend critically on a knowledge of the
nebular electron temperature.  However, the [\ion{O}{3}]
$\lambda$4363 is often weak and for two objects in our sample it is
suspect.  These cases and the solutions adopted are discussed below.

The spectrum of KISSR 666 gives an unusually high electron temperature,
over 20,000 K.  However, the [\ion{O}{3}] $\lambda$4363 lines in the individual 
Lick spectra (that were summed to produce the
final spectrum) appear noisy.  This is not surprising since the galaxy is very
faint with $m_B = 19.81$, much fainter than galaxies
traditionally studied for abundances.  In order to confirm this
unusual result, we decided to re-observe the object with
the Monolithic Mirror Telescope (MMT).  The larger aperture of the MMT
allowed for a cleaner spectrum which indicated a significantly lower 
electron temperature of 16,500 K.  Because the spectrum from the 
MMT had a higher signal-to-noise ratio, we adopted that temperature (Lee et
al. 2004) for the subsequent abundance analysis of the Lick spectrum.  

In a similar case, the spectrum of KISSB 23 appears to have a suspect [\ion{O}{3}]
$\lambda$4363 line.  The line observed in our Lick spectrum
is unphysically broad.  This odd
shape is repeated in all four of the individual spectra.  In addition,
the center of the putative [\ion{O}{3}] $\lambda$4363 line is offset
slightly in wavelength compared to its expected position.  We
re-observed this object on the MMT as well.  The MMT spectrum does not
reproduce the odd shape of the line, indicating that the Lick spectrum
is suspect.  We  hypothesize that scattered light in the KAST
spectrograph fell upon the location of the [\ion{O}{3}] $\lambda$4363
line giving it the observed broad appearance and leading to an
over-estimation of the line flux.
We again adopt the temperature calculated from the MMT
spectrum (14750 K, Lee et al. 2004).

\subsection{Calculating Metal Abundances}

We use the IRAF NEBULAR package (Shaw \& Dufour 1995) to calculate ionic 
abundances relative to hydrogen.  The input data tables include the 
densities and temperatures of each ionization zone, as calculated above, 
and the emission-line ratios measured previously (Table 6).  We calculate
the abundance of the following ions with respect to H$^+$:
O$^+$, O$^{++}$, N$^+$, S$^+$, S$^{++}$, Ne$^{++}$, and Ar$^{++}$. 

Ionization correction factors account for the additional ionization states 
present in the nebula that do not emit in the optical spectrum.  We use the 
prescription of Izotov et al. (1994) given by:

\begin{eqnarray}
ICF(N) &=&\frac{N}{N^+}=\frac{O}{O^+},\\
ICF(Ne) &=& \frac{Ne}{Ne^{++}}=\frac{O}{O^{++}},\\
ICF(S) & = & \frac{S}{S^+ + S^{++}} \nonumber\\
       & = & [0.013+x\{5.10+x[-12.78+x(14.77-6.11x)]\}]^{-1},\\
ICF(Ar)& = & \frac{Ar}{Ar^{++}} \nonumber\\
       & = & [0.15 + x(2.39 - 2.64x)]^{-1},\\
x & = & \frac{O^+}{O}.
\end{eqnarray}
In addition Izotov et al. offer a calculation for O$^{+++}$ given by:
\begin{equation}
\frac{O^{+++}}{O^{++}}=\frac{He^{++}}{He^{+}}.
\end{equation}
However, in the few cases where we observe He$^{++}$ in the spectra, the 
line is very noisy and we assume that the amount of O$^{+++}$ in the nebulae 
is negligible.

Using the ionic abundances and ionization correction factors given
above, we deduce the O/H, N/H, and Ne/H ratios for
each galaxy.  Measurements of the sulfur abundance ratios, S/H are
limited to the 6 galaxies in which we measure S$^{++}$.  We
measure Ar/H in 8 of 12 galaxies.  The abundance results are
presented in Table \ref{table:abund}.  (Note: preliminary  metallicities for
this sample of  galaxies were presented in Paper 1.  The metallicities 
reported here are updated and supersede the preliminary results.  
We plan to use the final results from this paper, in
conjunction with metallicity results from Lee et al. (2004) 
to update the metallicity -- line-ratio relationships given in Paper 1.)

%\placetable{table:abund}

We generate error estimates for the abundance data in Table
\ref{table:abund} in the following way.  We use the one sigma errors
in the oxygen 
line ratios to generate new T$_e$[OIII]$_{dev}$ measurements for each
system.  The error in the electron temperature is then given by,
\begin{equation}
\sigma (T_e[OIII]) = T_e[OIII] - T_e[OIII]_{dev}
\end{equation}

We then propagate this error estimate through equation 2 to obtain
$\sigma (T_e[OII])$.  We run the abundance programs with temperature offsets 
to generate a second abundance estimate for each element.  We take the
difference between the original ionic abundance estimate and the new estimate
as the error in the ionic abundance.  In order to account for the effects of 
errors in the observed line ratios, we add in quadrature the percentage error 
of the line ratio to the temperature-based error for each ion.  Errors in 
temperature range from 10\% to 4\% and result in final metallicity errors of 
0.04 -- 0.09 dex. A comparison of our metallicity result for the extremely
metal-deficient KISSR 1752 with Thuan et al. (1999) reveals good agreement.  
We find 12 + log(O/H)$ = 7.57 \pm 0.04$ while they give $7.60 \pm 0.01$.  In 
addition, for the three galaxies where we have an independent MMT spectrum we 
find good agreement in the metal abundances derived from the two spectra.

\section{Discussion of Metallicity Results}

\subsection{Abundances and Abundance Ratios}

The final abundance estimates are given in Table \ref{table:abund}.
It is standard to refer to the metallicity of a galaxy in units of 12
+ log(O/H).  In these units solar metallicity is 8.92 (Lambert
1978). (A more recent determination of the solar oxygen abundance by
Allende Prieto et al. (2001) gives a metallicity of 8.69.  We will however
continue to refer to the older solar abundance value so that we can more
easily compare with previous work.)

We find that all twelve galaxies studied here are metal poor with 12 $+$
log(O/H) $<$ 8.1.  The mean metallicty of the sample is 7.79 (1/13th solar),
and three galaxies are extremely metal deficient, approaching 12 $+$ log(O/H) 
$\sim$7.5, or $\sim1/25$ solar.  As mentioned above, the formal errors on
our abundances are somewhat higher than many previous studies, due to the
faintness of many of our sources.  The average error in the oxygen abundance
is 0.056 dex, and in all cases the errors are smaller than 0.1 dex.

%\placefigure{fig:prime}
\begin{figure*}[htp]
\epsfxsize=5.0in
\epsscale{1.0}
\vspace{-0.4in}
\plotone{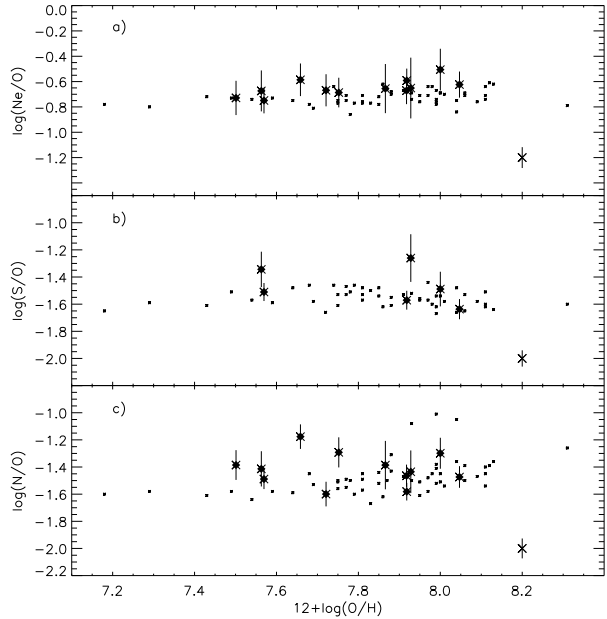}
\figcaption{Abundance ratios as a function of metallicity are plotted for Lick data
  (diamonds) and the Izotov and Thuan (1999) dataset (points).  Panel (a) shows the 
  neon to oxygen ratio, panel (b) is the sulfur to oxygen ratio, while panel (c) is
  the nitrogen to oxygen ratio.  Typical error bars for the Izotov and Thuan data
  are shown in the lower right of each plot.
  \label{fig:prime}}
\end{figure*}

Metal enrichment of galaxies occurs as a result of at
least two processes, ejection of enriched material during core
collapse and type Ia supernova, and matter loss through stellar
winds.  The $\alpha$ elements such as oxygen, sulfur,
neon and argon are produced in massive stars.  Because they
are created under the same conditions, the 
ratio of these elements should remain 
constant with increasing O/H ratios.  This has been demonstrated
empirically in Izotov and Thuan (1999), among others.  Panel (a) of 
Figure \ref{fig:prime} shows the Ne/O ratio as a
function of O/H for the Lick data (diamonds) and the Izotov and Thuan
dataset (points).  Panel (b) shows S/O.  In general the Lick data
agree with the Izotov and Thuan data, albeit with a somewhat higher scatter.
We find the mean log(Ne/O) $= -0.65 \pm 0.07$ while Izotov and Thuan gives 
log(Ne/O) $= -0.72 \pm 0.06$. Similarly for sulfur we find a mean log(S/O) 
$= -1.47 \pm 0.14$ with Izotov and Thuan reporting log(S/O) $= -1.56
\pm 0.06$.  Argon (not plotted) has similar behavior to neon and sulfur.  
We find a mean log(Ar/O) $= -2.20 \pm 0.13$, while Izotov and Thuan give
mean log(Ar/O) $= -2.26 \pm 0.09$.

The production of nitrogen is more complicated than the $\alpha$ elements
because nitrogen production can be enhanced by the CNO cycle in lower
mass stars.  However, for low
metallicity galaxies it has been found that N/O is constant for
increasing O/H (Izotov and Thuan 1999). In panel (c) of  Figure 
\ref{fig:prime} we plot the N/O ratio as a function of O/H again
comparing the Lick data (diamonds) to the Izotov and Thuan data
(points). The Lick data give a mean  log(N/O) $=
-1.42 \pm 0.12$ while Izotov and Thuan give  log(N/O) $=
-1.47 \pm 0.14$.  No significant nitrogen enhancements 
are noticeable in the figure, and we concur with van Zee et 
al. (1998) and Izotov \& Thuan (1999) that 
nitrogen is a primary element in low metallicity systems and only
becomes significantly enhanced with respect to oxygen at metallicities
higher than that of the galaxies contained in the Lick sample. 
  
Helium abundances, especially in low metallicity systems like the ones 
being studied here, are exciting to measure because they can place bounds 
on the primordial helium abundance.  While helium lines were detected in 
all twelve spectra, they were not high enough signal-to-noise to allow for 
a reasonably accurate determination of the helium abundance.  Higher quality 
spectra of these objects may prove interesting for future studies of helium 
abundances in low-metallicity systems.

\subsection{T$_e$ vs. $p_3$ Metallicities}

%\placefigure{fig:compmet}
\begin{figure*}[htp]
\epsfxsize=4.0in
\epsscale{0.8}
%\vspace{-0.5in}
\plotone{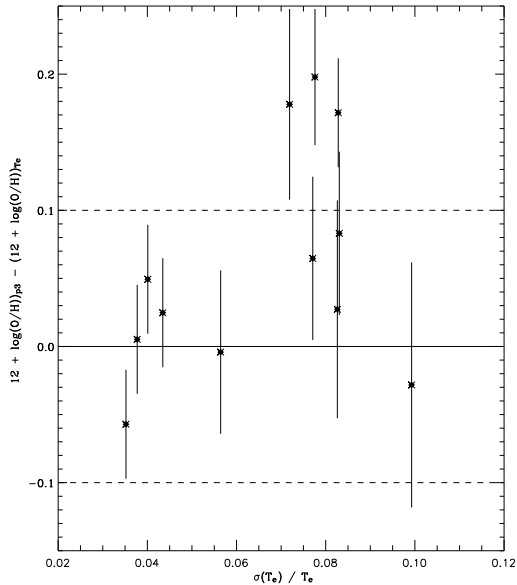}
%\vspace{-0.5in}
\figcaption{A comparison of metal abundances calculated with T$_e$ and
  $p_3$ methods.  The difference between the two methods is plotted
  against the factional error in the temperature 
  measurement.  
  \label{fig:compmet}}
\end{figure*}

Because the T$_e$ metal abundances are so dependent on the quality of the
[\ion{O}{3}]$\lambda4363$ line, we compare the T$_e$ abundance results 
with metallicities derived from the strong oxygen lines alone.  Recently, 
Pilyugin (2000, 2001) demonstrated a method for calculating metallicity 
from the strong oxygen lines [\ion{O}{2}]$\lambda\lambda3726,29$ and
[\ion{O}{3}]$\lambda\lambda4959,5007$.   
The Pilyugin method (hereafter the $p_3$ method) is an empirical
result that relates ratios of the strong oxygen lines and H$\beta$
with metallicities measured with the T$_e$ method.  He
finds that his method agrees with the T$_e$ method to within 0.1 dex for
low metallicity systems.  In fact he suggests that his 
method can be more accurate than the T$_e$ method when the
[\ion{O}{3}]$\lambda$4363 line is weak (Pilyugin 2001).  This claim is based
partly on data from the Hidalgo-Gamez and Oloffson (1998)
metallicity-luminosity work.  Hidalgo-Gamez and Oloffson found little
evidence for a metallicity-luminosity relation in their sample of
starbursting galaxies, where metallicity was measured with the T$_e$ method.  
When the metallicities were recalculated using the Pilyugin method, the data
closely follow the Skillman et al. (1989) metallicity-luminosity
relation.  This may indicate that the T$_e$ method is less reliable than the 
strong line method when [\ion{O}{3}]$\lambda$4363 is noisy.  

%\placefigure{fig:metlumte}
\begin{figure*}[htp]
\epsfxsize=5.0in
\epsscale{1.0}
\plotone{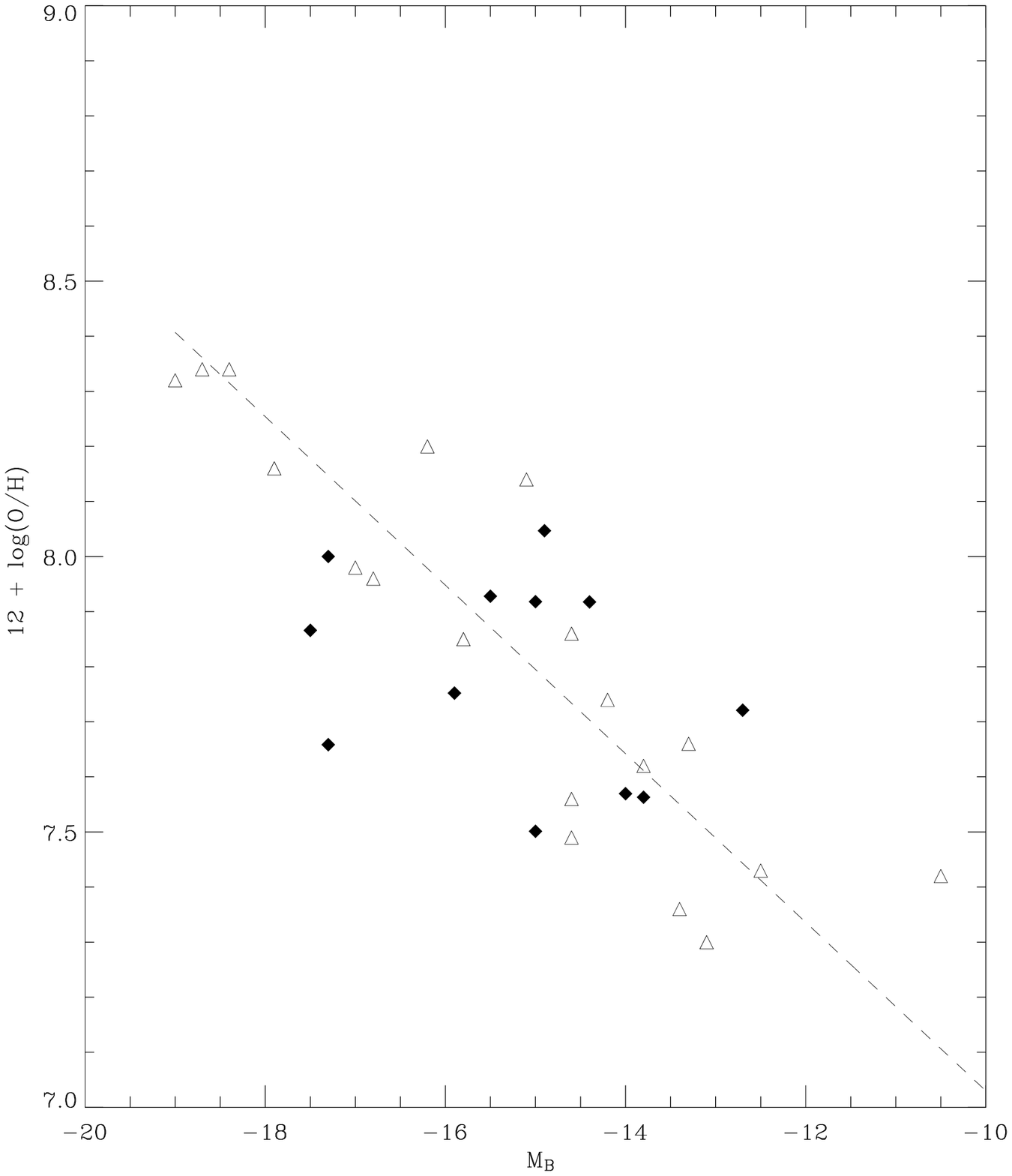}
\figcaption{A metallicity-luminosity relation.  The diamonds are
  data from this paper with metallicity calculated by the T$_e$
  method.  The triangles are data from Skillman et al. (1989). The
  dashed line is the fit to the Skillman et al. data.
  \label{fig:metlumte}}
\end{figure*}

%\placefigure{fig:metlump3}
\begin{figure*}[htp]
\epsfxsize=5.0in
\epsscale{1.0}
\plotone{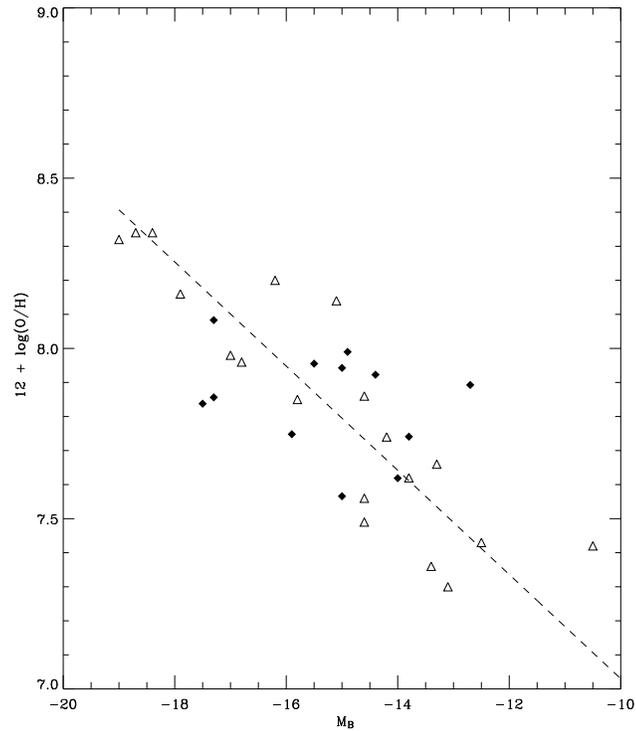}
\figcaption{The same as Figure \ref{fig:metlumte}, except that the metallicity
  for the Lick objects is now calculated with the $p_3$ method.  
  \label{fig:metlump3}}
\end{figure*}

We use the $p_3$ method to generate a second metallicity estimate for our galaxies.  
The results are included in Table \ref{table:abund}.  We find that nine of the twelve 
galaxies have $p_3$ abundances within 0.1 dex of their T$_e$ result.  The other three 
are within 0.2 dex.  In order to illustrate this result, we plot the difference 
between the T$_e$ and $p_3$ metallicity results vs $\sigma$(T$_e)/$T$_e$ in Figure 
\ref{fig:compmet}.  We see that all three deviant galaxies have 
$\sigma$(T$_e)/$T$_e > 0.07$ and tend to be noisier spectra. 

One of our most deviant objects is KISSB 23, which gives a $p_3$ metallicity 
0.17 dex higher than its $T_e$ result.  What is interesting in this case is that 
the excitation indicator [\ion{O}{3}]$\lambda$5007/[\ion{O}{2}]$\lambda\lambda$3726,29  
appears to be quite low, suggesting that this object may be past the peak of
its starburst phase, with the most massive stars already evolved off
the main sequence.  Objects like this are rare in most ELG surveys,
since post-peak objects are often fainter and have weaker emission
lines.  We have, however, found a similar object in the Skillman et
al. (1989) dataset.  Sextans A field 1 has a low excitation index
but a high temperature and low metallicity.  When we calculate a $p_3$
abundance for this object we find that $p_3$ gives an abundance 0.3
dex higher than the $T_e$ method.  This may be suggesting that $p_3$,
which has been calibrated with a dataset composed exclusively of high 
excitation objects, is not reliable for objects with lower excitation
spectra.

To further investigate whether $p_3$ or T$_e$ is more reliable, we
use Pilyugin's idea of looking for a tight metallicity-luminosity relation.  
In Figure \ref{fig:metlumte}, we plot T$_e$ abundances for our galaxies as 
filled diamonds.  For comparison, we include the Skillman et al. (1989) data 
shown as triangles. The fit to the Skillman data is shown as a dashed line.  
In Figure \ref{fig:metlump3} we plot the same, using our $p_3$ metallicities.  
We see that both the T$_e$ and $p_3$ results appear to mimic the Skillman
et al.  metallicity-luminosity relation.  The RMS scatter of the points about 
the Skillman fit for the T$_e$ plot is 0.25 while for the $p_3$ plot the 
$RMS = 0.24$.  In other words, there is no significant difference in the 
scatter about the Skillman et al. relation for the two methods.  One can 
improve the the T$_e$ plot by taking the $p_3$ results for KISSR 1490 and 
KISSR 1013, indicating that in a few instances the T$_e$ method may be 
underestimating the metallicity.  However, the position of galaxy KISSB 23 
on Figures \ref{fig:metlumte} and \ref{fig:metlump3} (M$_B$ $\sim-12.5$)  
indicates that the lower metallicity predicted by the T$_e$ method appears 
to be more consistent with the metallicity-luminosity relation than that
given by the $p_3$ method.  Recall that KISSB 23  appears to be  be atypical 
of hot young starbursts like those used in the sample on which the $p_3$ method 
is based. 
  
The $p_3$ method appears to work well at characterizing the metallicity of 
hot young starbursts such as those studied by Izotov and Thuan and most of 
the galaxies in this study.  In fact the $p_3$ results show that we may be 
underestimating the abundance of two galaxies where the 
[\ion{O}{3}]$\lambda$4363 line is noisy.  However it also appears that for 
low metallicity starbursts past their peak of star formation, the $p_3$ 
method may overestimate the metal abundance. In these rare cases a full 
nebular abundance approach may be the only reliable method for determining 
the metallicity.  

\section{Conclusions}
The KISS database contains many low
metallicity galaxies. These objects can be quickly identified 
from the original survey data from their estimated absolute blue
magnitudes and the 
metallicity-luminosity relation.  While a sample based on absolute
magnitude alone will  also contain a small percentage of  high
redshift AGN's,  the 
majority of systems selected this way are dwarf starburst galaxies. 
The selection criteria used
lead to the discovery of twelve metal-poor objects with 12 + log(O/H)
$<$ 8.1, three of which are extremely metal-poor objects with metallicities
approaching 1/25 solar.    

The Lick 3m telescope is a valuable instrument for obtaining
abundance-quality spectra of the dwarf systems in the KISS sample, producing 
metallicity measurements with formal errors of 0.04 - 0.09 dex. Using
these spectra, we confirm that in most cases the $p_3$ strong line abundances 
match well with the T$_e$ abundances
derived with the [\ion{O}{3}]$\lambda$4363 line.  In fact in two cases we see
evidence that the $p_3$ results may be an improvement on the T$_e$
results indicating  that we may be underestimating 
the errors in our metallicity measurements when the
[\ion{O}{3}]$\lambda$4363 line is noisy.   
We caution, however, that the $p_3$ method may break down 
for galaxies past the peak of their star burst.  

We calculated metal abundance ratios of O/H, N/H, 
Ne/H, S/H and Ar/H for twelve galaxies.  We found that the mean
metal abundance ratios with respect oxygen match those of
the Izotov et al. (1999) study.  

This current work is a first effort by the KISS collaboration to produce 
abundance-quality spectra of low-metallicity starburst galaxies.  As the
KISS collaboration generates more spectra of this type, we will be in
a position to contribute significantly to studies of star formation
mechanisms, star formation rates, and chemical evolution of 
the universe including primordial abundances. Because KISS samples a well
defined region of space we hope to be able to answer statistical questions 
such as how common are galaxies with metallicities 1/10 solar and 1/25 solar.  
We hope to provide further constraints on the rates of secondary 
metal production, and with additional follow-up work on the most metal-poor
KISS objects we may also be able to place bounds on the primordial helium
abundance of the universe.

\acknowledgements 

We gratefully acknowledge financial support for the KISS project from 
the NSF through awards NSF-AST-9553020 and NSF-AST-0071114 to JJS.  We 
also thank Wesleyan University for providing additional funding for the 
observing runs during which these spectral data were obtained.  
%Several useful suggestions by the anonymous referee helped to improve 
%the presentation of this paper.  
Finally, we wish to thank Dr. Stone and Dr. Gates of Lick Observatory for their 
expert assistance at the telescope when we obtained these observations.

% ******************************* Tables *************************************

\input{jmelbourne.table1.tex}

\input{jmelbourne.table2.tex}
\input{jmelbourne.table3.tex}
\input{jmelbourne.table4.tex}

\input{jmelbourne.table5.tex}
\input{jmelbourne.table6.tex}
\input{jmelbourne.table7.tex}

\end{document}

%% file: jmelbourne.table1.tex
\newpage
\rotate
\tabletypesize{\footnotesize}
\renewcommand{\arraystretch}{.6} 

\begin{deluxetable}{rrrrrlrrrrrrrrrrl}
\footnotesize
\tablewidth{0pt}
\tablecaption{Lick Spectroscopic Data: 30$^\circ$ Red Survey \label{table:lick30}}
\tablehead{
\colhead{KISSR} &
\colhead{Field} & 
\colhead{ID} & 
\colhead{Run} & 
\colhead{Q} & 
\colhead{$z$} & 
\colhead{$c_{H\beta}$} & 
\colhead{} & 
\colhead{EW} & 
\colhead{(\AA)} & 
\colhead{} & 
\colhead{H$\alpha$ flux\tablenotemark{1}} & 
\colhead{} & 
\colhead{Flux} & 
\colhead{ratios\tablenotemark{2}} & 
\colhead{} & 
\colhead{Type} \\[0.2ex]
\cline{8-11}
\cline{13-16}
\\[-1.3ex]
\colhead{} & 
\colhead{} & 
\colhead{} & 
\colhead{} & 
\colhead{} & 
\colhead{} & 
\colhead{} & 
\colhead{[OII]} & 
\colhead{H$\beta$} & 
\colhead{[OIII]} & 
\colhead{H$\alpha$} & 
\colhead{} & 
\colhead{$\frac{[O II]}{H\beta}$} & 
\colhead{$\frac{[O III]}{H\beta}$} & 
\colhead{$\frac{[N II]}{H\alpha}$} & 
\colhead{$\frac{[S II]}{H\alpha}$} &
\colhead{} 
}
\startdata
2 & F1215 & 4362 & 81 & 2 & 0.0301   & 1.15 & 11.08 & 2.71 & 0.91 & 21.40 & 84.900 & 0.7761 & -0.5319 & -0.3315 & -0.4771 & SBG \\ 
3 & F1215 & 4188 & 82 & 1 & 0.0268   & 0.66 & 29.15 & 8.54 & 5.97 & 56.55 & 73.410 & 0.7285 & -0.1692 & -0.5709 & -0.4073 & SBG \\ 
49 & F1220 & 380 & 81 & 1 & 0.0266   & 0.20 & 65.78 & 23.74 & 73.58 & 121.60 & 267.20 & 0.5365 & 0.5152 & -1.0990 & -0.6981 & SBG \\ 
85 & F1235 & 1574 & 83 & 1 & 0.0233  & 0.07 & 64.33 & 99.86 & 345.00 & 639.20 & 60.190 & 0.0111 & 0.5097 & -1.8038 & -1.2704 & SBG \\ 
97 & F1240 & 4022 & 81 & 1 & 0.0316  & 0.17 & 57.09 & 18.47 & 89.00 & 107.57 & 71.760 & 0.4933 & 0.6277 & -1.3369 & -0.8005 & SBG \\ 
101 & F1240 & 3527 & 82 & 2 & 0.0351 & 0.56 & 4.64  & 3.38 & 0.68 & 15.61 & 85.300 & 0.2404 & -0.6601 & -0.4650 & -0.5276 & SBG \\ 
114 & F1245 & 3999 & 83 & 3 & 0.0351 & 0.42 & 6.40  & 1.23 & 1.60 & 5.20 & 15.730 & 0.6690 & -0.0020 & -0.1992 & -0.4382 & SBG \\ 
120 & F1245 & 2181 & 81 & 1 & 0.0312 & 0.18 & 87.15 & 143.30 & 713.80 & 251.40 & 29.650 & 0.1666 & 0.6752 & -1.3674 & -1.0930 & SBG \\ 
158 & F1255 & 1524 & 83 & 2 & 0.0263 & 1.68 & 3.89 & 0.81 & 0.45 & 7.88 & 65.950 & 0.9572 & -0.2881 & -0.2788 & -0.6784 & SBG \\ 
396 & F1355 & 4898 & 82 & 1 & 0.0076 & 0.00 & 83.56 & 48.54 & 195.20 & 206.70 & 251.00 & 0.3959 & 0.6203 & -1.5312 & -0.9032 & SBG \\ 
404 & F1355 & 3016 & 81 & 1 & 0.0314 & -0.14 & 53.56 & 19.66 & 62.86 & 88.57 & 40.260 & 0.4593 & 0.4562 & -1.4893 & -0.7885 & SBG \\ 
492 & F1415 & 5463 & 81 & 3 & \nodata & \nodata & \nodata & \nodata & \nodata & \nodata & \nodata & \nodata & \nodata & \nodata & \nodata & Gal. \\ 
537 & F1420 & 2276 & 81 & 3 & \nodata & \nodata & \nodata & \nodata & \nodata & \nodata & \nodata & \nodata & \nodata & \nodata & \nodata & Gal. \\ 
666 & F1515 & 5729 & 82 & 1 & 0.0337 & 0.07 & 31.77 & 116.40 & 723.50 & 650.20 & 65.090 & -0.3724 & 0.7912 & -2.0472 & -1.5182 & SBG \\ 
675 & F1515 & 3377 & 81 & 1 & 0.0175 & 0.00 & 86.16 & 164.60 & 955.20 & 943.90 & 66.440 & 0.1036 & 0.7524 & -1.7055 & -1.0599 & SBG \\ 
757 & F1525 & 3026 & 81 & 2 & 0.0362 & 0.60 & 59.05 & 8.08 & 27.91 & 50.86 & 36.110 & 0.8725 & 0.5121 & -1.2739 & -0.4421 & SBG \\ 
785 & F1530 & 4705 & 81 & 2 & 0.0356 & -0.12 & 31.08 & 22.80 & 140.20 & 124.20 & 26.080 & 0.2875 & 0.7598 & -1.5646 & -0.9809 & SBG \\ 
840 & F1540 & 4806 & 83 & 2 & 0.0599 & 1.17 & 3.15 & 1.46 & \nodata & 9.24 & 44.050 & 0.5300 & \nodata & -0.3393 & -0.7707 & SBG \\ 
947 & F1600 & 6052 & 83 & 2 & 0.0776 & 1.40 & 11.56 & 1.92 & 2.85 & 15.44 & 28.820 & 1.1354 & 0.2690 & -0.4094 & -0.4336 & SBG \\ 
954 & F1600 & 4000 & 83 & 1 & \nodata & \nodata & \nodata & \nodata & \nodata & \nodata & \nodata & \nodata & \nodata & \nodata & \nodata & Star \\ 
1013 & F1615 & 4535 & 83 & 1 & 0.0249 & 0.03 & 47.74 & 33.77 & 91.02 & 91.10 & 61.450 & 0.3200 & 0.5916 & -1.2592 & -0.7327 & SBG \\ 
1021 & F1615 & 905 & 82 & 1 & 0.0085 & 0.13 & 64.18 & 11.52 & 35.25 & 59.00 & 50.250 & 0.7753 & 0.4676 & -1.3252 & -0.6081 & SBG \\ 
1024 & F1620 & 6657 & 81 & 3 & 0.0614 & \nodata & \nodata & \nodata & \nodata & 42.37 & 99.180 & \nodata & \nodata & -0.2302 & \nodata & SBG \\ 
1064 & F1635 & 110 & 82 & 2 & 0.1022 & -0.03 & 7.02 & 5.50 & \nodata & 9.07 & 17.130 & 0.0216 & \nodata & -0.3747 & -0.3351 & Gal. \\ 
1110 & F1655 & 11336 & 83 & 1 & 0.0693 & 1.69 & 14.57 & 2.34 & 2.63 & 23.42 & 87.580 & 1.0781 & 0.0275 & -0.2431 & -0.4851 & SBG \\ 
\tableline
\enddata

\tablenotetext{1}{In units of $10^{-16}$ ergs s$^{-1}$ cm$^{-2}$.}

\tablenotetext{2}{In logarithmic units.}

\end{deluxetable}

%% file: jmelbourne.table2.tex
\newpage
\rotate
\tabletypesize{\footnotesize}
\renewcommand{\arraystretch}{.6} 

\begin{deluxetable}{rrrrrlrrrrrrrrrrl}
\footnotesize
\tablewidth{0pt}
\tablecaption{Lick Spectroscopic Data: 43$^\circ$ Red Survey \label{table:lick43}}
\tablehead{
\colhead{KISSR} &
\colhead{Field} & 
\colhead{ID} & 
\colhead{Run} & 
\colhead{Q} & 
\colhead{$z$} & 
\colhead{$c_{H\beta}$} & 
\colhead{} & 
\colhead{EW} & 
\colhead{(\AA)} & 
\colhead{} & 
\colhead{H$\alpha$ flux\tablenotemark{1}} & 
\colhead{} & 
\colhead{Flux} & 
\colhead{ratios\tablenotemark{2}} & 
\colhead{} & 
\colhead{Type} \\[0.2ex]
\cline{8-11}
\cline{13-16}
\\[-1.3ex]
\colhead{} & 
\colhead{} & 
\colhead{} & 
\colhead{} & 
\colhead{} & 
\colhead{} & 
\colhead{} & 
\colhead{[OII]} & 
\colhead{H$\beta$} & 
\colhead{[OIII]} & 
\colhead{H$\alpha$} & 
\colhead{} & 
\colhead{$\frac{[O II]}{H\beta}$} & 
\colhead{$\frac{[O III]}{H\beta}$} & 
\colhead{$\frac{[N II]}{H\alpha}$} & 
\colhead{$\frac{[S II]}{H\alpha}$} &
\colhead{} 
}
\startdata
1194 & F1204 & 4743 & 161 & 1 & 0.0035 & 0.08 & 69.22 & 48.90 & 219.20 & 207.60 & 295.000 & 0.3114 & 0.7088 & -1.5057 & -0.9193 & SBG \\ 
1332 & F1234 & 6529 & 163 & 1 & 0.3190 & \nodata & 3.14 & 76.38 & 42.13 & \nodata & \nodata & \nodata & -0.3001 & \nodata & \nodata & QSO \\ 
1412 & F1259 & 2360 & 83 & 1 & 0.0595 & 1.28 & 69.03 & 9.46 & 16.11 & 73.04 & 76.192 & 1.1057 & 0.2421 & 0.3915 & 0.1221 & LIN \\ 
1490 & F1312 & 2143 & 161 & 1 & 0.0117 & -0.09 & 61.10 & 65.46 & 234.00 & 274.70 & 35.710 & 0.1789 & 0.5903 & -1.6063 & -0.9080 & SBG \\ 
1516 & F1316 & 5689 & 161 & 1 & 0.3277 & \nodata & 137.30 & 150.70 & 748.70 & \nodata & \nodata & 0.2991 & 0.7098 & \nodata & \nodata & SY2 \\ 
1751 & F1416 & 2866 & 83 & 1 & 0.3476 & 2.45 & 69.23 & 18.32 & 316.90 & \nodata & \nodata & \nodata & 1.1508 & \nodata & \nodata & SY2 \\ 
1752 & F1416 & 2273 & 163 & 1 & 0.0020 & -0.01 & 92.84 & 166.90 & 577.50 & 1130.00 & 779.377 & 0.0377 & 0.5532 & -1.8965 & -1.1726 & SBG \\ 
1778 & F1424 & 2639 & 161 & 1 & 0.0079 & 0.08 & 65.06 & 31.13 & 82.38 & 147.30 & 93.580 & 0.4663 & 0.4738 & -1.2948 & -0.6959 & SBG \\ 
1794 & F1428 & 2362 & 163 & 1 & 0.0083 & -0.11 & 41.91 & 29.49 & 89.21 & 105.00 & 36.029 & 0.3222 & 0.6115 & -1.4925 & -0.6858 & SBG \\ 
1845 & F1441 & 3484 & 161 & 1 & 0.0082 & 0.14 & 109.00 & 63.28 & 302.00 & 292.60 & 262.400 & 0.3932 & 0.7222 & -1.4294 & -0.8942 & SBG \\ 
1910 & F1507 & 2437 & 163 & 1 & 0.0180 & 0.32 & 44.99 & 12.31 & 37.10 & 100.70 & 15.917 & 0.7092 & 0.5213 & -1.1283 & -0.4574 & SBG \\ 
2097 & F1606 & 6404 & 82 & 1 & 0.3201 & \nodata & \nodata & 16.21 & 828.20 & \nodata & \nodata & \nodata & \nodata & \nodata & \nodata & SY2 \\ 
2151 & F1615 & 2010 & 163 & 3 & \nodata & \nodata & \nodata & \nodata & \nodata & \nodata & \nodata & \nodata & \nodata & \nodata & \nodata & Star \\ 
\tableline
\enddata

\tablenotetext{1}{In units of $10^{-16}$ ergs s$^{-1}$ cm$^{-2}$.}

\tablenotetext{2}{In logarithmic units.}

\end{deluxetable}

%% file: jmelbourne.table3.tex
\newpage
\rotate
\tabletypesize{\footnotesize}
\renewcommand{\arraystretch}{.6} 

\begin{deluxetable}{rrrrrlrrrrrrrrrrl}
\footnotesize
\tablewidth{0pt}
\tablecaption{Lick Spectroscopic Data: 30$^\circ$ Blue Survey \label{table:lickblue}}
\tablehead{
\colhead{KISSB\tablenotemark{1}} &
\colhead{Field} & 
\colhead{ID} & 
\colhead{Run} & 
\colhead{Q} & 
\colhead{$z$} & 
\colhead{$c_{H\beta}$} & 
\colhead{} & 
\colhead{EW} & 
\colhead{(\AA)} & 
\colhead{} & 
\colhead{H$\alpha$ flux\tablenotemark{2}} & 
\colhead{} & 
\colhead{Flux} & 
\colhead{ratios\tablenotemark{3}} & 
\colhead{} & 
\colhead{Type} \\[0.2ex]
\cline{8-11}
\cline{13-16}
\\[-1.3ex]
\colhead{} & 
\colhead{} & 
\colhead{} & 
\colhead{} & 
\colhead{} & 
\colhead{} & 
\colhead{} & 
\colhead{[OII]} & 
\colhead{H$\beta$} & 
\colhead{[OIII]} & 
\colhead{H$\alpha$} & 
\colhead{} & 
\colhead{$\frac{[O II]}{H\beta}$} & 
\colhead{$\frac{[O III]}{H\beta}$} & 
\colhead{$\frac{[N II]}{H\alpha}$} & 
\colhead{$\frac{[S II]}{H\alpha}$} &
\colhead{} 
}
\startdata
23 & F0935 & 158 & 161 & 1 & 0.0018 & 0.14 & 46.59 & 31.45 & 65.67 & 162.60 & 207.400 & 0.4301 & 0.3233 & -1.5246 & -0.8826 & SBG \\
94 & F1220 & 380 & 81 & 1 & 0.0266 & 0.20 & 65.78 & 23.74 & 73.58 & 121.60 & 267.20 & 0.5365 & 0.5152 & -1.0990 & -0.6981 & SBG \\ 
112 & F1255 & 1524 & 83 & 2 & 0.0263 & 1.68 & 3.89 & 0.81 & 0.45 & 7.88 & 65.950 & 0.9572 & -0.2881 & -0.2788 & -0.6784 & SBG \\ 
145 & F1355 & 4898 & 82 & 1 & 0.0076 & 0.00 & 83.56 & 48.54 & 195.20 & 206.70 & 251.000 & 0.3959 & 0.6203 & -1.5312 & -0.9032 & SBG \\
186 & F1515 & 5729 & 82 & 1 & 0.0337 & 0.07 & 31.77 & 116.40 & 723.50 & 650.20 & 65.090 & -0.3724 & 0.7912 & -2.0472 & -1.5182 & SBG \\  
187 & F1515 & 3377 & 81 & 1 & 0.0175 & 0.00 & 86.16 & 164.60 & 955.20 & 943.90 & 66.440 & 0.1036 & 0.7524 & -1.7055 & -1.0599 & SBG \\ 
194 & F1530 & 4705 & 81 & 2 & 0.0356 & -0.12 & 31.08 & 22.80 & 140.20 & 124.20 & 26.080 & 0.2875 & 0.7598 & -1.5646 & -0.9809 & SBG \\ 
211 & F1615 & 4535 & 83 & 1 & 0.0249 & 0.03 & 47.74 & 33.77 & 91.02 & 91.10 & 61.450 & 0.3200 & 0.5916 & -1.2592 & -0.7327 & SBG \\ 
\tableline
\enddata

\tablenotetext{1}{KISSB 94 = KISSR 49; KISSB 112 = KISSR 158; KISSB 145 = KISSR 396; KISSB 186 = KISSR 666; KISSB 187 = KISSR 675; KISSB 194 = KISSR 785; KISSB 211 = KISSR 1013.}
\tablenotetext{2}{In units of $10^{-16}$ ergs s$^{-1}$ cm$^{-2}$.}
\tablenotetext{3}{In logarithmic units.}

\end{deluxetable}

%% file: jmelbourne.table4.tex
\newpage
\rotate
\tabletypesize{\footnotesize}
\renewcommand{\arraystretch}{.6} 

\begin{deluxetable}{rrrrrc}
\footnotesize
\tablewidth{0pt}
\tablecaption{Properties Of Galaxies With Abundance-Quality Spectra \label{table:galprop}}
\tablehead{
\colhead{KISSR} &
\colhead{m$_B$} &
\colhead{B$-$V} &
\colhead{M$_B$} &
\colhead{Velocity} & 
\colhead{12 + log(O/H)$_{T_e}$} \\
&&&& \colhead{[km/s]} &
}
\startdata

    49  & 17.83  &  0.58 & -17.42 & 7872 &  8.00\\   
    85  & 19.91  &  0.11 & -15.04 & 6939 &  7.50\\   
   396  & 17.49  &  0.32 & -15.03 & 2317 &  7.92\\   
   666  & 19.83  &  0.39 & -15.96 & 9954 &  7.75\\  
   675  & 16.87  &  0.33 & -17.52 & 5314 &  7.87\\   
  1013  & 17.79  &  0.37 & -17.44 & 7541 &  7.66\\   
  1194  & 16.40  &  0.50 & -14.47 & 1088 &  7.92\\   
  1490  & 19.58  &  0.50 & -13.90 & 3559 &  7.56\\   
  1752  & 15.61  &  0.21 & -14.35 &  721 &  7.57\\  
  1778  & 17.09  &  0.43 & -15.56 & 2476 &  7.93\\    
  1845  & 17.83  &  0.66 & -14.95 & 2590 &  8.05\\    
 23\tablenotemark{1} & 16.32 &  0.22 & -12.46 &  412 & 7.72\\

\tableline
\enddata

\tablenotetext{1}{KISSB}

\end{deluxetable}

%% file: jmelbourne.table5.tex
\newpage
\rotate
\tabletypesize{\scriptsize}
\renewcommand{\arraystretch}{.6} 

\begin{deluxetable}{lcrrrrrrrrrrrr}
\footnotesize
\tablewidth{0pt}
\tablecaption{Observed Line Ratios with respect to H$\beta$ \label{table:lineratios}}
\tablehead{
\\[-1.1ex]
&&&&&&& KISSR &&&&&& KISSB \\[0.2ex]
\cline{3-13}
\\[-1.2ex]
\colhead{Ion} & \colhead{$\lambda$ } &
\colhead{49} &   \colhead{85} &  \colhead{396} &
\colhead{666} &  \colhead{675} &  
\colhead{1013}&    \colhead{1194} &  \colhead{1490} &
\colhead{1752} &  \colhead{1778} & \colhead{1845} & \colhead{23}
 }
\startdata
$[$O II$]$ B  &         3728 & 
   2.980  &    0.977  &    2.488  &    0.404  &    1.269  &    2.050  &    1.940  &    1.510  &    1.091  &    2.679  &    2.238  &    2.352  \\
 &   &  
 $\pm$  0.145  &  $\pm$  0.039  &  $\pm$  0.070  &  $\pm$  0.021  &  $\pm$  0.073  &  $\pm$  0.074  &  $\pm$  0.066  &  $\pm$  0.065  &  $\pm$  0.040  &  $\pm$  0.097  &  $\pm$  0.076  &  $\pm$  0.081  \\ 
 H 10  &         3798 &   &  &  &  &  &  &  &  &  &  &    0.027  &  \\ 
 &   &    &  &  &  &  &  &  &  &  &  &  $\pm$  0.002  &  \\ 
 H 9  &         3836 & 
 &  &  &  &  &  &  &  &    0.057  &  &    0.055  &  \\ 
 &   &    &  &  &  &  &  &  &  &  $\pm$  0.003  &  &  $\pm$  0.003  &  \\ 
$$[$$Ne III$$]$$  &         3869 & 
   0.329  &    0.236  &    0.404  &    0.498  &    0.485  &    0.406  &    0.399  &    0.347  &    0.257  &    0.229  &    0.433  &    0.159  \\
 &   &  
 $\pm$  0.019  &  $\pm$  0.014  &  $\pm$  0.013  &  $\pm$  0.022  &  $\pm$  0.033  &  $\pm$  0.020  &  $\pm$  0.015  &  $\pm$  0.020  &  $\pm$  0.010  &  $\pm$  0.012  &  $\pm$  0.016  &  $\pm$  0.008  \\ 
 He I+H 8  &         3889 & 
 &    0.166  &    0.180  &    0.226  &    0.158  &    0.093  &    0.144  &    0.151  &    0.188  &    0.161  &    0.158  &    0.151  \\
 &   &  
 &  $\pm$  0.012  &  $\pm$  0.007  &  $\pm$  0.014  &  $\pm$  0.019  &  $\pm$  0.011  &  $\pm$  0.006  &  $\pm$  0.013  &  $\pm$  0.008  &  $\pm$  0.010  &  $\pm$  0.007  &  $\pm$  0.008  \\ 
$[$Ne III$]$+H$\epsilon$  &         3970 & 
 &    0.189  &    0.238  &    0.273  &  &  &    0.210  &    0.204  &    0.225  &    0.158  &    0.245  &    0.123  \\
 &   &  
 &  $\pm$  0.012  &  $\pm$  0.009  &  $\pm$  0.015  &  &  &  $\pm$  0.008  &  $\pm$  0.014  &  $\pm$  0.009  &  $\pm$  0.010  &  $\pm$  0.009  &  $\pm$  0.007  \\ 
 H$\delta$  &         4102 & 
   0.237  &    0.216  &    0.267  &    0.220  &    0.245  &    0.180  &    0.262  &    0.266  &    0.276  &    0.231  &    0.259  &    0.243  \\
 &   &  
 $\pm$  0.014  &  $\pm$  0.013  &  $\pm$  0.009  &  $\pm$  0.013  &  $\pm$  0.022  &  $\pm$  0.013  &  $\pm$  0.010  &  $\pm$  0.016  &  $\pm$  0.011  &  $\pm$  0.012  &  $\pm$  0.010  &  $\pm$  0.010  \\ 
 H$\gamma$  &         4340 & 
   0.467  &    0.490  &    0.484  &    0.453  &    0.480  &    0.358  &    0.462  &    0.503  &    0.472  &    0.437  &    0.464  &    0.421  \\
 &   &  
 $\pm$  0.024  &  $\pm$  0.021  &  $\pm$  0.015  &  $\pm$  0.019  &  $\pm$  0.032  &  $\pm$  0.018  &  $\pm$  0.016  &  $\pm$  0.024  &  $\pm$  0.018  &  $\pm$  0.018  &  $\pm$  0.016  &  $\pm$  0.016  \\ 
 $[$O III$]$  &         4363 & 
   0.042  &    0.086  &    0.069  &    0.228  &    0.110  &    0.106  &    0.090  &    0.120  &    0.091  &    0.039  &    0.073  &    0.057  \\
 &   &  
 $\pm$  0.006  &  $\pm$  0.008  &  $\pm$  0.005  &  $\pm$  0.013  &  $\pm$  0.017  &  $\pm$  0.011  &  $\pm$  0.005  &  $\pm$  0.011  &  $\pm$  0.004  &  $\pm$  0.007  &  $\pm$  0.003  &  $\pm$  0.005  \\ 
 He I  &         4472 & 
   0.019  &    0.052  &    0.031  &    0.037  &  &  &    0.038  &  &    0.035  &  &    0.037  &    0.022  \\
 &   &  
 $\pm$  0.006  &  $\pm$  0.007  &  $\pm$  0.004  &  $\pm$  0.007  &  &  &  $\pm$  0.003  &  &  $\pm$  0.002  &  &  $\pm$  0.002  &  $\pm$  0.004  \\ 
 He II  &         4687 & 
 &  &  &  &  &  &  &  &    0.026  &  &    0.015  &  \\ 
 &   &    &  &  &  &  &  &  &  &  $\pm$  0.002  &  &  $\pm$  0.002  &  \\ 
 H$\beta$  &         4861 & 
   1.000  &    1.000  &    1.000  &    1.000  &    1.000  &    1.000  &    1.000  &    1.000  &    1.000  &    1.000  &    1.000  &    1.000  \\
 &   &  
 $\pm$  0.049  &  $\pm$  0.036  &  $\pm$  0.028  &  $\pm$  0.036  &  $\pm$  0.056  &  $\pm$  0.037  &  $\pm$  0.034  &  $\pm$  0.042  &  $\pm$  0.037  &  $\pm$  0.037  &  $\pm$  0.034  &  $\pm$  0.035  \\ 
 He I  &         4922 &   &  &  &  &  &  &  &  &  &  &    0.012  &  \\ 
 &   &    &  &  &  &  &  &  &  &  &  &  $\pm$  0.001  &  \\ 
 $[$O III$]$  &         4959 & 
   1.077  &    1.048  &    1.381  &    2.085  &    1.863  &    1.257  &    1.699  &    1.330  &    1.159  &    0.968  &    1.776  &    0.644  \\
 &   &  
 $\pm$  0.053  &  $\pm$  0.038  &  $\pm$  0.039  &  $\pm$  0.069  &  $\pm$  0.100  &  $\pm$  0.045  &  $\pm$  0.057  &  $\pm$  0.054  &  $\pm$  0.043  &  $\pm$  0.036  &  $\pm$  0.059  &  $\pm$  0.023  \\ 
 $[$O III$]$  &         5007 & 
   3.320  &    3.248  &    4.171  &    6.212  &    5.655  &    3.912  &    5.141  &    3.893  &    3.574  &    2.900  &    5.325  &    2.061  \\
 &   &  
 $\pm$  0.160  &  $\pm$  0.108  &  $\pm$  0.114  &  $\pm$  0.198  &  $\pm$  0.295  &  $\pm$  0.130  &  $\pm$  0.170  &  $\pm$  0.150  &  $\pm$  0.130  &  $\pm$  0.103  &  $\pm$  0.176  &  $\pm$  0.070  \\ 
 He I  &         5876 & 
   0.109  &    0.095  &    0.110  &    0.131  &    0.127  &    0.084  &  &    0.112  &  &    0.116  &    0.121  &  \\ 
 &   &  
 $\pm$  0.012  &  $\pm$  0.008  &  $\pm$  0.004  &  $\pm$  0.008  &  $\pm$  0.025  &  $\pm$  0.008  &  &  $\pm$  0.018  &  &  $\pm$  0.012  &  $\pm$  0.008  &  \\ 
 $[$O I$]$  &         6300 & 
   0.082  &    0.054  &    0.053  &  &  &    0.078  &    0.045  &    0.070  &    0.024  &    0.059  &    0.033  &    0.023  \\
 &   &  
 $\pm$  0.009  &  $\pm$  0.006  &  $\pm$  0.003  &  &  &  $\pm$  0.008  &  $\pm$  0.004  &  $\pm$  0.012  &  $\pm$  0.002  &  $\pm$  0.007  &  $\pm$  0.003  &  $\pm$  0.003  \\ 
 $[$S III$]$  &         6312 & 
   0.025  &  &    0.019  &  &  &  &    0.022  &    0.034  &    0.017  &    0.036  &    0.018  &  \\ 
 &   &  
 $\pm$  0.005  &  &  $\pm$  0.002  &  &  &  &  $\pm$  0.002  &  $\pm$  0.008  &  $\pm$  0.001  &  $\pm$  0.005  &  $\pm$  0.002  &  \\ 
 $[$O I$]$  &         6364 & 
 &  &    0.016  &  &  &  &    0.017  &  &  &  &    0.014  &  \\ 
 &   &  
 &  &  $\pm$  0.002  &  &  &  &  $\pm$  0.002  &  &  &  &  $\pm$  0.002  &  \\ 
 $[$N II$]$  &         6548 & 
   0.096  &    0.021  &    0.023  &  &    0.030  &    0.056  &    0.028  &  &    0.018  &    0.043  &    0.035  &    0.025  \\
 &   &  
 $\pm$  0.011  &  $\pm$  0.005  &  $\pm$  0.003  &  &  $\pm$  0.011  &  $\pm$  0.007  &  $\pm$  0.003  &  &  $\pm$  0.001  &  $\pm$  0.006  &  $\pm$  0.003  &  $\pm$  0.003  \\ 
 H$\alpha$  &         6563 & 
   3.278  &    2.913  &    2.801  &    2.923  &    2.789  &    2.821  &    2.962  &    2.564  &    2.761  &    2.894  &    3.126  &    2.976  \\
 &   &  
 $\pm$  0.322  &  $\pm$  0.136  &  $\pm$  0.063  &  $\pm$  0.104  &  $\pm$  0.499  &  $\pm$  0.130  &  $\pm$  0.183  &  $\pm$  0.371  &  $\pm$  0.123  &  $\pm$  0.284  &  $\pm$  0.206  &  $\pm$  0.211  \\ 
 $[$N II$]$  &         6583 & 
   0.262  &    0.046  &    0.082  &    0.026  &    0.055  &    0.155  &    0.093  &    0.063  &    0.035  &    0.147  &    0.116  &    0.089  \\
 &   &  
 $\pm$  0.026  &  $\pm$  0.006  &  $\pm$  0.003  &  $\pm$  0.004  &  $\pm$  0.014  &  $\pm$  0.011  &  $\pm$  0.006  &  $\pm$  0.011  &  $\pm$  0.002  &  $\pm$  0.015  &  $\pm$  0.008  &  $\pm$  0.007  \\ 
 He I  &         6678 & 
   0.038  &  &    0.024  &  &    0.060  &  &    0.035  &  &    0.039  &    0.026  &    0.033  &    0.025  \\
 &   &  
 $\pm$  0.006  &  &  $\pm$  0.003  &  &  $\pm$  0.015  &  &  $\pm$  0.003  &  &  $\pm$  0.002  &  $\pm$  0.004  &  $\pm$  0.003  &  $\pm$  0.003  \\ 
 $[$S II$]$  &         6717 & 
   0.380  &    0.077  &    0.203  &    0.056  &    0.114  &    0.274  &    0.206  &    0.188  &    0.105  &    0.336  &    0.233  &    0.235  \\
 &   &  
 $\pm$  0.038  &  $\pm$  0.007  &  $\pm$  0.006  &  $\pm$  0.005  &  $\pm$  0.023  &  $\pm$  0.016  &  $\pm$  0.013  &  $\pm$  0.028  &  $\pm$  0.005  &  $\pm$  0.034  &  $\pm$  0.016  &  $\pm$  0.017  \\ 
 $[$S II$]$  &         6731 & 
   0.287  &    0.080  &    0.147  &    0.033  &    0.129  &    0.249  &    0.153  &    0.129  &    0.081  &    0.250  &    0.171  &    0.160  \\
 &   &  
 $\pm$  0.029  &  $\pm$  0.007  &  $\pm$  0.005  &  $\pm$  0.004  &  $\pm$  0.026  &  $\pm$  0.015  &  $\pm$  0.010  &  $\pm$  0.020  &  $\pm$  0.004  &  $\pm$  0.025  &  $\pm$  0.012  &  $\pm$  0.012  \\ 
 He I  &         7065 & 
 &  &    0.027  &  &  &  &    0.017  &  &    0.023  &  &    0.039  &    0.017  \\
 &   &  
 &  &  $\pm$  0.003  &  &  &  &  $\pm$  0.002  &  &  $\pm$  0.002  &  &  $\pm$  0.003  &  $\pm$  0.002  \\ 
 $[$Ar III$]$  &         7136 & 
   0.062  &  &    0.068  &  &  &  &    0.113  &    0.078  &    0.043  &    0.100  &    0.105  &    0.041  \\
 &   &  
 $\pm$  0.008  &  &  $\pm$  0.003  &  &  &  &  $\pm$  0.007  &  $\pm$  0.013  &  $\pm$  0.002  &  $\pm$  0.011  &  $\pm$  0.007  &  $\pm$  0.004  \\ 
 $[$O II$]$  &         7319 & 
 &  &    0.045  &  &  &  &    0.031  &  &  &  &    0.029  &  \\ 
 &   &  
 &  &  $\pm$  0.003  &  &  &  &  $\pm$  0.003  &  &  &  &  $\pm$  0.002  &  \\

\tableline 
\enddata

\end{deluxetable}

%\tablenotetext{a}{This is the running number for the galaxy as listed
%in Salzer et al. 2001.}
%\tablenotetext{b}{This is the KISSB number as listed in Salzer et al. 2002a.}
%
% H 10  &         3798 &   &  &  &  &  &  &  &  &  &  &    0.027  &  \\ 
% &   &    &  &  &  &  &  &  &  &  &  &  $\pm$  0.002  &  \\ 
%
% H 9  &         3836 & 
%  &  &  &  &  &  &  &  &    0.057  &  &    0.055  &  \\ 
% &   &    &  &  &  &  &  &  &  &  $\pm$  0.003  &  &  $\pm$  0.003  &  \\ 
%
% He I  &         4026 &     0.012  &  &  &  &  &  &  &  &  &  &  &  \\ 
% &   &    $\pm$  0.007  &  &  &  &  &  &  &  &  &  &  &  \\ 
%
% $$[$$S II$$]$$ B  &         4073 &   &  &  &    0.033  &  &  &  &  &  &  &  &  \\ 
% &   &    &  &  &  $\pm$  0.008  &  &  &  &  &  &  &  &  \\ 
%
% He I  &         4922 &   &  &  &  &  &  &  &  &  &  &    0.012  &  \\ 
% &   &    &  &  &  &  &  &  &  &  &  &  $\pm$  0.001  &  \\ 
%
% $$[$$O II$$]$$ B  &         7325 &   &  &  &  &  &  &  &  &    0.030  &  &  &  \\ 
% &   &    &  &  &  &  &  &  &  &  $\pm$  0.002  &  &  &  \\ 

%% file: jmelbourne.table6.tex
\newpage
\rotate
\tabletypesize{\scriptsize}
\renewcommand{\arraystretch}{.6} 

\begin{deluxetable}{lcrrrrrrrrrrrr}
\footnotesize
\tablewidth{0pt}
\tablecaption{Corrected Line Ratios with respect to H$\beta$ \label{table:lineratioscor}}
\tablehead{
\\[-1.1ex]
&&&&&&& KISSR &&&&&& KISSB \\[0.2ex]
\cline{3-13}
\\[-1.2ex]
\colhead{Ion} & \colhead{$\lambda$} &
\colhead{49} &   \colhead{85} &  \colhead{396} &
\colhead{666} &  \colhead{675} &  
\colhead{1013}&    \colhead{1194} &  \colhead{1490} &
\colhead{1752} &  \colhead{1778} & \colhead{1845} & \colhead{23}
 }
\startdata
 $[$O II$]$ B  &         3728 & 
   3.440  &    1.026  &    2.488  &    0.424  &    1.269  &    2.089  &    2.048  &    1.510  &    1.091  &    2.796  &    2.429  &    2.536  \\
 &   &  
 $\pm$  0.201 &  $\pm$  0.061 &  $\pm$  0.070 &  $\pm$  0.027 &  $\pm$  0.073 &  $\pm$  0.120 &  $\pm$  0.132 &  $\pm$  0.065 &  $\pm$  0.040 &  $\pm$  0.236 &  $\pm$  0.162 &  $\pm$  0.177 \\ 
 H 10  &         3798 &   &  &  &  &  &  &  &  &  &  &    0.030  &  \\ 
 &   &    &  &  &  &  &  &  &  &  &  &  $\pm$  0.003 &  \\ 
 H 9  &         3836 & 
 &  &  &  &  &  &  &  &    0.057  &  &    0.059  &  \\ 
 &   &    &  &  &  &  &  &  &  &  $\pm$  0.003 &  &  $\pm$  0.004 &  \\ 
 $[$Ne III$]$  &         3869 & 
   0.373  &    0.246  &    0.404  &    0.519  &    0.485  &    0.413  &    0.418  &    0.347  &    0.257  &    0.237  &    0.465  &    0.169  \\
 &   &  
 $\pm$  0.023 &  $\pm$  0.017 &  $\pm$  0.013 &  $\pm$  0.029 &  $\pm$  0.033 &  $\pm$  0.026 &  $\pm$  0.025 &  $\pm$  0.020 &  $\pm$  0.010 &  $\pm$  0.020 &  $\pm$  0.028 &  $\pm$  0.012 \\ 
 He I+H 8  &         3889 & 
 &    0.173  &    0.180  &    0.235  &    0.158  &    0.094  &    0.150  &    0.151  &    0.188  &    0.167  &    0.170  &    0.161  \\
 &   &  
 &  $\pm$  0.014 &  $\pm$  0.007 &  $\pm$  0.016 &  $\pm$  0.019 &  $\pm$  0.012 &  $\pm$  0.010 &  $\pm$  0.013 &  $\pm$  0.008 &  $\pm$  0.015 &  $\pm$  0.011 &  $\pm$  0.012 \\ 
 $[$Ne III$]$+H$\epsilon$  &         3970 & 
 &    0.196  &    0.238  &    0.283  &  &  &    0.219  &    0.204  &    0.225  &    0.164  &    0.261  &    0.130  \\
 &   &  
 &  $\pm$  0.014 &  $\pm$  0.009 &  $\pm$  0.017 &  &  &  $\pm$  0.013 &  $\pm$  0.014 &  $\pm$  0.009 &  $\pm$  0.014 &  $\pm$  0.015 &  $\pm$  0.009 \\ 
 H$\delta$  &         4102 & 
   0.260  &    0.223  &    0.267  &    0.227  &    0.245  &    0.182  &    0.271  &    0.266  &    0.276  &    0.237  &    0.273  &    0.255  \\
 &   &  
 $\pm$  0.016 &  $\pm$  0.014 &  $\pm$  0.009 &  $\pm$  0.014 &  $\pm$  0.022 &  $\pm$  0.014 &  $\pm$  0.014 &  $\pm$  0.016 &  $\pm$  0.011 &  $\pm$  0.017 &  $\pm$  0.014 &  $\pm$  0.015 \\ 
 H$\gamma$  &         4340 & 
   0.497  &    0.500  &    0.484  &    0.462  &    0.480  &    0.361  &    0.473  &    0.503  &    0.472  &    0.445  &    0.481  &    0.434  \\
 &   &  
 $\pm$  0.027 &  $\pm$  0.023 &  $\pm$  0.015 &  $\pm$  0.021 &  $\pm$  0.032 &  $\pm$  0.019 &  $\pm$  0.020 &  $\pm$  0.024 &  $\pm$  0.018 &  $\pm$  0.024 &  $\pm$  0.021 &  $\pm$  0.020 \\ 
 $[$O III$]$  &         4363 & 
   0.045  &    0.088  &    0.069  &    0.233  &    0.110  &    0.107  &    0.092  &    0.120  &    0.091  &    0.040  &    0.076  &    0.059  \\
 &   &  
 $\pm$  0.007 &  $\pm$  0.009 &  $\pm$  0.005 &  $\pm$  0.013 &  $\pm$  0.017 &  $\pm$  0.011 &  $\pm$  0.005 &  $\pm$  0.011 &  $\pm$  0.004 &  $\pm$  0.007 &  $\pm$  0.004 &  $\pm$  0.005 \\ 
 He I  &         4472 & 
   0.020  &    0.053  &    0.031  &    0.038  &  &  &    0.039  &  &    0.035  &  &    0.038  &    0.023  \\
 &   &  
 $\pm$  0.006 &  $\pm$  0.007 &  $\pm$  0.004 &  $\pm$  0.007 &  &  &  $\pm$  0.003 &  &  $\pm$  0.002 &  &  $\pm$  0.002 &  $\pm$  0.005 \\ 
 He II  &         4687 & 
 &  &  &  &  &  &  &  &    0.026  &  &    0.015  &  \\ 
 &   &    &  &  &  &  &  &  &  &  $\pm$  0.002 &  &  $\pm$  0.002 &  \\ 
 H$\beta$  &         4861 & 
   1.000  &    1.000  &    1.000  &    1.000  &    1.000  &    1.000  &    1.000  &    1.000  &    1.000  &    1.000  &    1.000  &    1.000  \\
 &   &  
 $\pm$  0.049 &  $\pm$  0.036 &  $\pm$  0.028 &  $\pm$  0.036 &  $\pm$  0.056 &  $\pm$  0.037 &  $\pm$  0.034 &  $\pm$  0.042 &  $\pm$  0.037 &  $\pm$  0.037 &  $\pm$  0.034 &  $\pm$  0.035 \\ 
 He I  &         4922 &   &  &  &  &  &  &  &  &  &  &    0.012  &  \\ 
 &   &    &  &  &  &  &  &  &  &  &  &  $\pm$  0.001 &  \\ 
 $[$O III$]$  &         4959 & 
   1.067  &    1.045  &    1.381  &    2.079  &    1.863  &    1.255  &    1.693  &    1.330  &    1.159  &    0.966  &    1.767  &    0.641  \\
 &   &  
 $\pm$  0.053 &  $\pm$  0.038 &  $\pm$  0.039 &  $\pm$  0.069 &  $\pm$  0.100 &  $\pm$  0.045 &  $\pm$  0.057 &  $\pm$  0.054 &  $\pm$  0.043 &  $\pm$  0.036 &  $\pm$  0.059 &  $\pm$  0.023 \\ 
 $[$O III$]$  &         5007 & 
   3.275  &    3.233  &    4.171  &    6.184  &    5.655  &    3.905  &    5.115  &    3.893  &    3.574  &    2.888  &    5.283  &    2.046  \\
 &   &  
 $\pm$  0.158 &  $\pm$  0.108 &  $\pm$  0.114 &  $\pm$  0.198 &  $\pm$  0.295 &  $\pm$  0.130 &  $\pm$  0.171 &  $\pm$  0.150 &  $\pm$  0.130 &  $\pm$  0.104 &  $\pm$  0.177 &  $\pm$  0.070 \\ 
 He I  &         5876 & 
   0.099  &    0.092  &    0.110  &    0.126  &    0.127  &    0.083  &  &    0.112  &  &    0.113  &    0.115  &  \\ 
 &   &  
 $\pm$  0.011 &  $\pm$  0.008 &  $\pm$  0.004 &  $\pm$  0.008 &  $\pm$  0.025 &  $\pm$  0.009 &  &  $\pm$  0.018 &  &  $\pm$  0.013 &  $\pm$  0.009 &  \\ 
 $[$O I$]$  &         6300 & 
   0.072  &    0.052  &    0.053  &  &  &    0.077  &    0.043  &    0.070  &    0.024  &    0.057  &    0.031  &    0.022  \\
 &   &  
 $\pm$  0.009 &  $\pm$  0.007 &  $\pm$  0.003 &  &  &  $\pm$  0.008 &  $\pm$  0.004 &  $\pm$  0.012 &  $\pm$  0.002 &  $\pm$  0.008 &  $\pm$  0.003 &  $\pm$  0.003 \\ 
 $[$S III$]$  &         6312 & 
   0.022  &  &    0.019  &  &  &  &    0.021  &    0.034  &    0.017  &    0.035  &    0.017  &  \\ 
 &   &  
 $\pm$  0.005 &  &  $\pm$  0.002 &  &  &  &  $\pm$  0.002 &  $\pm$  0.008 &  $\pm$  0.001 &  $\pm$  0.006 &  $\pm$  0.002 &  \\ 
 $[$O I$]$  &         6364 & 
 &  &    0.016  &  &  &  &    0.017  &  &  &  &    0.013  &  \\ 
 &   &  
 &  &  $\pm$  0.002 &  &  &  &  $\pm$  0.002 &  &  &  &  $\pm$  0.002 &  \\ 
 $[$N II$]$  &         6548 & 
   0.083  &    0.020  &    0.023  &  &    0.030  &    0.055  &    0.027  &  &    0.018  &    0.041  &    0.032  &    0.023  \\
 &   &  
 $\pm$  0.010 &  $\pm$  0.005 &  $\pm$  0.003 &  &  $\pm$  0.011 &  $\pm$  0.008 &  $\pm$  0.003 &  &  $\pm$  0.001 &  $\pm$  0.006 &  $\pm$  0.003 &  $\pm$  0.003 \\ 
 H$\alpha$  &         6563 & 
   2.813  &    2.767  &    2.801  &    2.778  &    2.789  &    2.765  &    2.796  &    2.564  &    2.761  &    2.765  &    2.865  &    2.747  \\
 &   &  
 $\pm$  0.293 &  $\pm$  0.182 &  $\pm$  0.063 &  $\pm$  0.150 &  $\pm$  0.499 &  $\pm$  0.184 &  $\pm$  0.238 &  $\pm$  0.371 &  $\pm$  0.123 &  $\pm$  0.352 &  $\pm$  0.257 &  $\pm$  0.264 \\ 
 $[$N II$]$  &         6583 & 
   0.224  &    0.043  &    0.082  &    0.025  &    0.055  &    0.152  &    0.087  &    0.063  &    0.035  &    0.140  &    0.107  &    0.082  \\
 &   &  
 $\pm$  0.024 &  $\pm$  0.006 &  $\pm$  0.003 &  $\pm$  0.004 &  $\pm$  0.014 &  $\pm$  0.013 &  $\pm$  0.008 &  $\pm$  0.011 &  $\pm$  0.002 &  $\pm$  0.019 &  $\pm$  0.010 &  $\pm$  0.008 \\ 
 He I  &         6678 & 
   0.032  &  &    0.024  &  &    0.060  &  &    0.032  &  &    0.039  &    0.025  &    0.030  &    0.023  \\
 &   &  
 $\pm$  0.005 &  &  $\pm$  0.003 &  &  $\pm$  0.015 &  &  $\pm$  0.003 &  &  $\pm$  0.002 &  $\pm$  0.005 &  $\pm$  0.003 &  $\pm$  0.003 \\ 
 $[$S II$]$  &         6717 & 
   0.321  &    0.073  &    0.203  &    0.053  &    0.114  &    0.268  &    0.194  &    0.188  &    0.105  &    0.320  &    0.211  &    0.215  \\
 &   &  
 $\pm$  0.034 &  $\pm$  0.008 &  $\pm$  0.006 &  $\pm$  0.005 &  $\pm$  0.023 &  $\pm$  0.021 &  $\pm$  0.018 &  $\pm$  0.028 &  $\pm$  0.005 &  $\pm$  0.043 &  $\pm$  0.020 &  $\pm$  0.022 \\ 
 $[$S II$]$  &         6731 & 
   0.243  &    0.075  &    0.147  &    0.031  &    0.129  &    0.244  &    0.143  &    0.129  &    0.081  &    0.238  &    0.155  &    0.146  \\
 &   &  
 $\pm$  0.026 &  $\pm$  0.008 &  $\pm$  0.005 &  $\pm$  0.004  &  $\pm$  0.026 &  $\pm$  0.019 &  $\pm$  0.013 &  $\pm$  0.020 &  $\pm$  0.004 &  $\pm$  0.032 &  $\pm$  0.015 &  $\pm$  0.015 \\ 
 He I  &         7065 & 
 &  &    0.027  &  &  &  &    0.016  &  &    0.023  &  &    0.035  &    0.015  \\
 &   &  
 &  &  $\pm$  0.003 &  &  &  &  $\pm$  0.002 &  &  $\pm$  0.002 &  &  $\pm$  0.004 &  $\pm$  0.002 \\ 
 $[$Ar III$]$  &         7136 & 
   0.052  &  &    0.068  &  &  &  &    0.105  &    0.078  &    0.043  &    0.095  &    0.094  &    0.037  \\
 &   &  
 $\pm$  0.007 &  &  $\pm$  0.003 &  &  &  &  $\pm$  0.010 &  $\pm$  0.013 &  $\pm$  0.002 &  $\pm$  0.014 &  $\pm$  0.010 &  $\pm$  0.004 \\ 
 $[$O II$]$  &         7319 & 
 &  &    0.045  &  &  &  &    0.029  &  &  &  &    0.026  &  \\ 
 &   &  
 &  &  $\pm$  0.003 &  &  &  &  $\pm$  0.003 &  &  &  &  $\pm$  0.003 &  \\ 
\hline
$c_{H\beta}$ & &  0.198 &0.067 & 0.00& 0.066& 0.00& 0.026& 0.075&
 0.00& 0.00& 0.059& 0.113& 0.104 \\

\tableline 
\enddata

\end{deluxetable}

%\tablenotetext{a}{This is the running number for the galaxy as listed
%in Salzer et al. 2001.}
%\tablenotetext{b}{This is the KISSB number as listed in Salzer et al. 2002a.}
%
% H 10  &         3798 &   &  &  &  &  &  &  &  &  &  &    0.030  &  \\ 
% &   &    &  &  &  &  &  &  &  &  &  &  $\pm$  0.003 &  \\ 
%
% H 9  &         3836 & 
%  &  &  &  &  &  &  &  &    0.057  &  &    0.059  &  \\ 
% &   &    &  &  &  &  &  &  &  &  $\pm$  0.003 &  &  $\pm$  0.004 &  \\ 
%
% He I  &         4026 &     0.014  &  &  &  &  &  &  &  &  &  &  &  \\ 
% &   &    $\pm$  0.009 &  &  &  &  &  &  &  &  &  &  &  \\ 
%
% $$[$$S II$$]$$ B  &         4073 &   &  &  &    0.034  &  &  &  &  &  &  &  &  \\ 
% &   &    &  &  &  $\pm$  0.008 &  &  &  &  &  &  &  &  \\ 
%
% He I  &         4922 &   &  &  &  &  &  &  &  &  &  &    0.012  &  \\ 
% &   &    &  &  &  &  &  &  &  &  &  &  $\pm$  0.001 &  \\ 
%
% $$[$$O II$$]$$ B  &         7325 &   &  &  &  &  &  &  &  &    0.030  &  &  &  \\ 
% &   &    &  &  &  &  &  &  &  &  $\pm$  0.002 &  &  &  \\ 

%% file: jmelbourne.table7.tex
\newpage
\rotate
\tabletypesize{\footnotesize}
\renewcommand{\arraystretch}{.6} 

\begin{deluxetable}{lrrrrrrrrrrrr}
\footnotesize
\tablewidth{0pt}
\tablecaption{Derived Metal Abundances. \label{table:abund}}
\tablehead{
\\[-1.1ex]
&&&&&& KISSR &&&&&& KISSB \\[0.2ex]
\cline{2-12}
\\[-1.2ex]
\colhead{Parameter}  &
\colhead{49} &   \colhead{85} &  \colhead{396} &
\colhead{666} &  \colhead{675} &  
\colhead{1013}&    \colhead{1194} &  \colhead{1490} &
\colhead{1752} &  \colhead{1778} & \colhead{1845} & \colhead{23}
}

\startdata
N$_e$ [1/cm$^3$] &           96&	       100&	        28&	       100&	       100&	       100&	        47&	       100&	       114&	        74&	        52&	       100\\
T$_e$ [OII] [K] &         12759 &         14707 &         13223 &         14248 &         13669 &         14733 &         13448 &         15079 &         14489 &         12763 &         12857 &         13530\\
   &		     $\pm$       518&	$\pm$       466&	$\pm$       270&	$\pm$       345&	$\pm$       614&	$\pm$       469&	$\pm$       234&	$\pm$       430&	$\pm$       244&	$\pm$       516&	$\pm$       210&	$\pm$       514\\
   T$_e$ [OIII] [K] &      13029&         17861&         14034&         16565&         15081&         17939&         14553&         18999&         17231&         13037&         13235&         14745\\
   &		     $\pm$      1082&	$\pm$      1377&	$\pm$       610&	$\pm$       935&	$\pm$      1497&	$\pm$      1392&	$\pm$       549&	$\pm$      1366&	$\pm$       691&	$\pm$      1077&	$\pm$       446&	$\pm$      1221\\
\hline
O$^+$/H$^+$ [$\times10^{-5}$] &         4.97 &        0.92 &        3.03 &        0.42 &        1.45 &        1.87 &        2.42 &        1.26 &        1.02 &        4.01 &        3.32 &        3.00 \\
    &   $\pm$      0.62 &  $\pm$      0.08 &  $\pm$      0.20 &  $\pm$      0.03 &  $\pm$      0.19 &  $\pm$      0.17 &  $\pm$      0.16 &  $\pm$      0.13 &  $\pm$      0.06 &  $\pm$      0.68 &  $\pm$      0.21 &  $\pm$      0.34 \\
O$^{++}$/H$^+$ [$\times10^{-5}$] &         5.03 &        2.25 &        5.25 &        5.23 &        5.90 &        2.68 &        5.85 &        2.40 &        2.69 &        4.46 &        7.82 &        2.26 \\
    &   $\pm$      1.18 &  $\pm$      0.40 &  $\pm$      0.68 &  $\pm$      0.72 &  $\pm$      1.46 &  $\pm$      0.48 &  $\pm$      0.84 &  $\pm$      0.53 &  $\pm$      0.37 &  $\pm$      1.48 &  $\pm$      1.08 &  $\pm$      0.34 \\
O/H    [$\times 10^{-5}$] &        10.00 &        3.17 &        8.28 &        5.65 &        7.34 &        4.55 &        8.27 &        3.66 &        3.71 &        8.47 &       11.14 &        5.26 \\
    &   $\pm$      1.33 &  $\pm$      0.41 &  $\pm$      0.71 &  $\pm$      0.72 &  $\pm$      1.47 &  $\pm$      0.51 &  $\pm$      0.85 &  $\pm$      0.55 &  $\pm$      0.38 &  $\pm$      1.63 &  $\pm$      1.10 &  $\pm$      0.48 \\
\hline
N$^+$/H$^+$ [$\times 10^{-6}$] &         2.50 &        0.38 &        0.79 &        0.22 &        0.60 &        1.25 &        0.83 &        0.49 &        0.33 &        1.47 &        1.12 &        0.75 \\
    &   $\pm$      0.33 &  $\pm$      0.06 &  $\pm$      0.04 &  $\pm$      0.04 &  $\pm$      0.16 &  $\pm$      0.13 &  $\pm$      0.08 &  $\pm$      0.09 &  $\pm$      0.02 &  $\pm$      0.25 &  $\pm$      0.11 &  $\pm$      0.09 \\
 N/H  [$\times 10^{-6}$] &         5.03 &        1.31 &        2.17 &        2.88 &        3.02 &        3.04 &        2.84 &        1.41 &        1.20 &        3.11 &        3.74 &        1.32 \\
    &   $\pm$      1.14 &  $\pm$      0.28 &  $\pm$      0.27 &  $\pm$      0.64 &  $\pm$      1.08 &  $\pm$      0.53 &  $\pm$      0.45 &  $\pm$      0.36 &  $\pm$      0.16 &  $\pm$      0.95 &  $\pm$      0.58 &  $\pm$      0.25 \\
\hline
 Ne$^{++}$/H$^+$ [$\times 10^{-6}$] &        15.70 &        4.19 &       13.45 &       10.76 &       13.04 &        6.96 &       12.48 &        5.08 &        4.80 &        9.96 &       18.61 &        4.85 \\
    &   $\pm$      3.62 &  $\pm$      0.75 &  $\pm$      1.67 &  $\pm$      1.56 &  $\pm$      3.14 &  $\pm$      1.24 &  $\pm$      1.70 &  $\pm$      1.07 &  $\pm$      0.58 &  $\pm$      3.47 &  $\pm$      2.47 &  $\pm$      1.04 \\
 Ne/H [$\times 10^{-6}$] &        31.22 &        5.92 &       21.20 &       11.63 &       16.25 &       11.82 &       17.63 &        7.75 &        6.61 &       18.92 &       26.50 &       11.28 \\
    &   $\pm$     11.09 &  $\pm$      1.68 &  $\pm$      4.22 &  $\pm$      2.76 &  $\pm$      6.48 &  $\pm$      3.25 &  $\pm$      3.94 &  $\pm$      2.64 &  $\pm$      1.39 &  $\pm$      9.80 &  $\pm$      5.72 &  $\pm$      3.13 \\
\hline
S$^+$/H$^+$     [$\times10^{-7} $] &         7.57 &        1.52 &        4.31 &        0.91 &        2.85 &        5.23 &        4.03 &        3.11 &        1.96 &        7.45 &        4.79 &        4.32 \\
    &   $\pm$      0.98 &  $\pm$      0.18 &  $\pm$      0.21 &  $\pm$      0.09 &  $\pm$      0.61 &  $\pm$      0.49 &  $\pm$      0.40 &  $\pm$      0.49 &  $\pm$      0.11 &  $\pm$      1.22 &  $\pm$      0.48 &  $\pm$      0.53 \\
 S$^{++}$/H$^+ $     [$\times10^{-7} $] &        19.42 &      &       13.08 &      &      &      &      &        9.66 &        6.29 &       30.86 &       14.24 &      \\
    &   $\pm$      6.35 &      &  $\pm$      2.16 &      &      &      &      &  $\pm$      3.12 &  $\pm$      0.86 &  $\pm$     12.36 &  $\pm$      2.45 &      \\
S/H     [$\times10^{-6} $] &         3.25 &      &        2.22 &         &        &      &      &        1.66 &        1.15 &        4.65 &        2.57 &      \\
    &   $\pm$      0.85 &      &  $\pm$      0.30 &  & &      &      &  $\pm$      0.43 &  $\pm$      0.13 &  $\pm$      1.66 &  $\pm$      0.36 &      \\
\hline
 Ar$^{++}$/H$^+ $     [$\times10^{-7} $] &         2.72 &      &        3.09 &      &      &      &        4.47 &        2.15 &        1.38 &        4.96 &        4.77 &        1.54 \\
    &   $\pm$      0.53 &      &  $\pm$      0.27 &      &      &      &  $\pm$      0.53 &  $\pm$      0.44 &  $\pm$      0.11 &  $\pm$      1.18 &  $\pm$      0.60 &  $\pm$      0.26 \\
Ar/H     [$\times10^{-7} $] &         3.97 &      &        4.61 &      &      &      &        7.18 &        3.26 &        2.27 &        7.19 &        7.60 &        2.35 \\
    &   $\pm$      0.93 &      &  $\pm$      0.48 &      &      &      &  $\pm$      0.95 &  $\pm$      0.74 &  $\pm$      0.22 &  $\pm$      2.13 &  $\pm$      1.05 &  $\pm$      0.49 \\
\hline
12 + log(O/H)$_{T_e}$ 	 &         8.00 &        7.50 &        7.92 &        7.75 &        7.87 &        7.66 &        7.92 &        7.56 &        7.57 &        7.93 &        8.05 &        7.72 \\
   		 &   $\pm$      0.06 &  $\pm$      0.06 &  $\pm$      0.04 &  $\pm$      0.06 &  $\pm$      0.09 &  $\pm$      0.05 &  $\pm$      0.04 &  $\pm$      0.07 &  $\pm$      0.04 &  $\pm$      0.08 &  $\pm$      0.04 &  $\pm$      0.04 \\
12 + log(O/H)$_{p_3}$ 	 & 8.08 & 7.57& 7.94 & 7.75 & 7.84 & 7.86 &
   		 7.92& 7.74 & 7.61 & 7.95 & 7.99 & 7.89 \\
12 + log(O/H)$_{T_e}$ MMT & &7.61 && 7.76& &&&&&&& 7.65\\ 

\tableline 
\enddata

\end{deluxetable}

%\tablenotetext{a}{This is the running number for the galaxy as listed
%in Salzer et al. 2001.}
%\tablenotetext{b}{This is the KISSB number as listed in Salzer et al. 2002a.}